\pdfoutput=1
\documentclass[a4paper,11pt]{article}
\usepackage{amsmath,amssymb,amsfonts}
\usepackage[multiple,stable]{footmisc}
\usepackage[amsmath,amsthm,thmmarks]{ntheorem}
\allowdisplaybreaks[4]
\usepackage[noconfig]{refstyle}
\usepackage[dvipsnames]{xcolor}
\usepackage[hyperfootnotes=false, linktocpage=true, colorlinks, citecolor=blue, linkcolor=blue, urlcolor=Maroon]{hyperref}
\usepackage{array,tabularx,booktabs,geometry,subfigure,graphicx,longtable}
\usepackage[bf]{caption}
\usepackage{tikz}
\usepackage{cite}
\usepackage{dsfont}
\usetikzlibrary{shapes,arrows}
\tikzstyle{block} = [rectangle, draw, text width=7em, text centered, rounded corners, minimum height=3em]

\newtheorem{theorem}{Theorem}[section]

\let\eqref=\relax
\newref{eq}{name={eq.~},Name={Eq.~},names={eqs.~},Names={Eqs.~},rngtxt={-},refcmd=(\ref{#1})}
\newref{f}{name={footnote~},Name={Footnote~},names={footnotes~},Names={Footnotes~}}
\newref{app}{name={appendix~},Name={Appendix~},names={appendixes~},Names={Appendixes~}}
\newref{tab}{name={table~},Name={Table~},names={tables~},Names={Tables~}}
\newref{ch}{name={chapter~},Name={Chapter~},names={chapters~},Names={Chapters~}}
\newref{sec}{name={section~},Name={Section~},names={sections~},Names={Sections~}}
\newref{fig}{name={figure~},Name={Figure~},names={figures~},Names={Figures~}}
\numberwithin{equation}{section}

\newcommand{\be}{\begin{equation}}
\newcommand{\ee}{\end{equation}}
\newcommand{\bea}{\begin{equation}\begin{aligned}}	
\newcommand{\eea}{\end{aligned}\end{equation}}		

\newcommand{\tr}{\mathrm{tr}}

\newcommand{\iddots}{\mathinner{\mkern2mu\raise1pt\hbox{.}\mkern2mu \raise4pt\hbox{.}\mkern2mu\raise7pt\hbox{.}\mkern1mu}}

\newcommand{\CS}{{\rm CS}}
\DeclareMathOperator{\diag}{diag}

\providecommand{\id}{\leavevmode\hbox{\small$\mathrm{1}$\kern-3.8pt\normalsize$\mathrm{1}$}}
\def\fnote#1#2{\begingroup\def\thefootnote{#1}\footnote{#2}
     \addtocounter{footnote}{-1}\endgroup}

\begin{document}

\vspace{1cm}

\title{
       {\Large \bf Chern-Simons Invariants and Heterotic Superpotentials}}

\vspace{2cm}

\author{
Lara~B.~Anderson${}^{1}$,
James~Gray${}^{1}$,
Andre~Lukas${}^{2}$,
Juntao~Wang${}^{1}$
}
\date{}
\maketitle
\begin{center} {\small ${}^1${\it Department of Physics, 
Robeson Hall, Virginia Tech \\ Blacksburg, VA 24061, U.S.A.}\\[0.2cm]
       ${}^2${\it Rudolf Peierls Centre for Theoretical Physics, Oxford
       University,\\
       $~~~~~$ 1 Keble Road, Oxford, OX1 3NP, U.K.}}\\

\fnote{}{jamesgray@vt.edu}
\fnote{}{lara.anderson@vt.edu}
\fnote{}{lukas@physics.ox.ac.uk} 
\fnote{}{wjunt88@vt.edu}

\end{center}

\begin{abstract}
\noindent
The superpotential in four-dimensional heterotic effective theories contains terms arising from holomorphic Chern-Simons invariants associated to the gauge and tangent bundles of the compactification geometry. These effects are crucial for a number of key features of the theory, including vacuum stability and moduli stabilization. Despite their importance, few tools exist in the literature to compute such effects in a given heterotic vacuum. In this work we present new techniques to explicitly determine holomorphic Chern-Simons invariants in heterotic string compactifications. The key technical ingredient in our computations are real bundle morphisms between the gauge and tangent bundles. We find that there are large classes of examples, beyond the standard embedding, where the Chern-Simons superpotential vanishes. We also provide explicit examples for non-flat bundles where it is non-vanishing and fractionally quantized, generalizing previous results for Wilson lines.
\end{abstract}

\thispagestyle{empty}
\setcounter{page}{0}
\newpage

\tableofcontents

\section{Introduction}
The literature on model building in smooth Calabi-Yau compactifications of heterotic string theory stretches back nearly 40 years. In the early seminal work on the subject, efforts were focussed on the case of the ``standard embedding" where the gauge bundle was taken to be the holomorphic tangent bundle of the Calabi-Yau manifold \cite{Candelas:1985en,gsw,Greene:1986bm,Greene:1986jb,Braun:2011ni}. In more recent years, advances in the technology used to describe these compactifications has lead to the construction of heterotic standard models with the exact charged spectrum of the MSSM \cite{Bouchard:2005ag,Braun:2011ni,Braun:2005ux,Braun:2005bw,Braun:2005zv,Anderson:2009mh,Anderson:2011ns,Anderson:2012yf}. Most of this model building progress has been achieved by branching out to more general situations where the gauge fields and the spin connection are connections on different holomorphic vector bundles. Despite the sophisticated constructions that lead to carefully chosen charged particle spectra, it has generally been the case that these compactifications give rise to only marginally stable vacua. Nevertheless, this work has been motivated by the hope that at least some of the general lessons learned in this decade-spanning and extensive effort will carry over to more realistic situations where all of the moduli are stabilized.\\[2mm]
Compared to string model building focused on particle physics properties, the subject of moduli stabilization is at a much less advanced stage of development in heterotic theories. Although a large literature on this topic does exist, several pieces of the low energy effective theory that would be required for a full analysis of the vacuum space of the theory are still unknown. These include, for example, the K\"ahler potential, in the case of non-standard embeddings, for fields such as matter and bundle moduli as an explicit function of the ${\cal N}=1$ degrees of freedom \cite{Gray:2003vw,Candelas:2016usb,McOrist:2016cfl,Blesneag:2018ygh,Candelas:2018lib}. In this work \emph{our goal is to compute another such missing component of the theory -- namely the vacuum contribution to the superpotential that appears due to the presence of the gauge bundle in heterotic compactifications}. If this quantity is non-vanishing it can potentially destabilize the model, in the absence of other effects. This superpotential due to the heterotic gauge bundle is also a crucial ingredient in moduli stabilization scenarios and so its computation is of great importance.\\[2mm]
In heterotic compactifications on smooth Calabi-Yau three-folds, we typically consider a gauge bundle $V$ over some Calabi-Yau manifold $X$ with tangent bundle $TX$. The Bianchi identity of the ten-dimensional theory, in the absence of five-branes is\footnote{Throughout this paper we will define `$\tr$' to include a factor of $\frac{1}{8 \pi^2}$ to avoid unnecessary cluttering of the formulae with numerical factors.},
\begin{eqnarray} \label{gbi}
dH= \alpha' \left( \tr (R\wedge R) - \tr (F\wedge F) \right) \;.
\end{eqnarray}
This implies that the Neveu-Schwarz three-form field strength can be written, at least locally, as
\begin{equation} \label{Hdef}
  H=H_0 + \alpha' \left(\omega_3(\omega)-\omega_3(A) \right)\;,\quad
  \omega_3(A) = {\rm tr}(dA \wedge A + \frac{2}{3} A^3)\;,\quad
    \omega_3(\omega) = {\rm tr}(d\omega \wedge \omega + \frac{2}{3} \omega^3)\; .
\end{equation}  
Here, $H_0$ is a closed contribution to the field strength that obeys an integer flux quantization condition and it can be locally written as $H_0=dB$, with the two-form field $B$. Further, we have introduced the gauge connection $A$ on $V$ and the spin connection $\omega$ on $TX$, along with their respective Chern-Simons forms  $\omega_3(A)$ and  $\omega_3(\omega)$.

Given such a form for the field strength $H$, the Gukov-Vafa-Witten superpotential \cite{Gukov:1999ya} of the four dimensional effective theory can be written as
\begin{equation} \label{GVW}
W = \int_X \left(H + i\, dJ \right)\wedge\Omega =\int_X H_0\wedge \Omega+\alpha'\,\CS_{\rm phys}(A,\omega)+i\int_XdJ\wedge\Omega\; ,
\end{equation}
where $\Omega$ is the holomorphic $(3,0)$ form on the Calabi-Yau three-fold. The physics literature usually defines the Chern-Simons contribution to this superpotential as
\begin{eqnarray} \label{phcs}
\CS_{\rm phys}(A,\omega) = \int_X \tr \left( \omega_3(A) - \omega_3 (\omega)\right) \wedge \Omega \; .
\end{eqnarray}
It is often tacitly assumed that this contribution vanishes in vacuum. In general, however,  there is no reason for this to be the case and it must be computed explicitly, even to verify the existence of simple forms of marginally stable Minkowski vacua\footnote{We thank E.~Witten for pointing this out to us and suggesting we consider this issue in the context of the work \cite{Anderson:2011ns,Anderson:2012yf}.}. It is possible that many of the heterotic standard models in the literature are not, in fact, associated to Minkowski vacua unless other contributions to (\ref{GVW}) are included. To date, only one of the  models with an exact MSSM spectrum mentioned above is known to have a vanishing Chern-Simons contribution to the superpotential~\cite{Braun:2011ni}. The gauge bundle, in this model, is a holomorphic deformation of the tangent bundle and in such instances, as we will discuss, the vanishing of (\ref{phcs}) is guaranteed.

Unfortunately, very few techniques exist in either the physics or mathematics literature for explicitly computing the value of (\ref{phcs}) and only a handful of simple examples have been studied. This is frustrating given its importance for heterotic model building and moduli stabilization. The goal of the present work is to improve the tools required for calculating Chern-Simons contributions to the superpotential in relevant heterotic models. In particular, our primary results are the following.
\begin{itemize}
\item We develop new computational tools to efficiently calculate the vacuum value of (\ref{phcs}) in explicit heterotic string compactifications.
\item We construct new and non-trivial examples of consistent heterotic compactifications in which the Chern-Simons contribution to the vacuum superpotential can be exactly determined. These include cases with vanishing as well as non-vanishing and fractional Chern-Simons contributions.
\end{itemize}
It should be noted that Wilson line contributions to the superpotential (\ref{phcs}) have been frequently considered in the literature \cite{Conrad:2000tk,Gukov:2003cy,Cicoli:2013rwa,Apruzzi:2014dza}. We emphasize that this is {\it not} what we are doing here. We are interested in all contributions to (\ref{phcs}), including those from the non-flat bundles. It is this quantity which is of relevance for concrete models, since heterotic compactifications on Calabi-Yau three-folds necessarily require a gauge bundle with non-vanishing curvature. Other recent papers considering this contribution to the superpotential even for non-flat bundles include \cite{Jockers:2009ti} which utilizes mirror symmetry and \cite{Ashmore:2018ybe} which explores deformations of the Hull-Strominger system \cite{Strominger:1986uh,Hull:1986kz}.\\[2mm]
To understand the physical consequences for non-vanishing Chern-Simons contributions to the vacuum potential, it is important to keep in mind that the effect being described here could, of course, be cancelled by the other contributions to the superpotential (\ref{GVW}). However, the Chern-Simons term is somewhat different in nature to the effects from $H_0$ and $dJ$. The contributions to (\ref{GVW}) from $H_0$ are associated with quantized quantities, which means that the corresponding terms in the superpotential are determined by a set of integers. The same is not necessarily true for the Chern-Simons contribution (\ref{phcs}). The Chern-Simons term is determined, as we will discuss, by a set of $2(h^{2,1}(X)+1)$ numbers which may be fractional. The contribution from $dJ$, by contrast,  vanishes for any torsion-free background and thus represents a highly non-trivial modification of the background geometry if present. Such a modification would have to be taken into account in other aspects of the dimensional reduction, for example in model building work. If the Chern-Simons contribution (\ref{phcs}) is non-zero it will typically be large so that, in the absence of other effects, it will destabilize the theory. On the other hand, any credible scenario for moduli stabilization which may, for example, include additional non-perturbative effects, must include the Chern-Simons contribution.
Fractional Chern-Simons contributions obtained from Wilson lines have already been used in some moduli stabilization scenarios \cite{Gukov:2003cy,Cicoli:2013rwa}.\\[2mm]
\emph{In any eventuality, it is important to understand what values the Chern-Simons term (\ref{phcs}) takes in compactifications of heterotic string theory, and it is this question that we will try to address in the rest of this paper.} \\[2mm]
In the next section we review the proper formulation of holomorphic Chern-Simons terms in heterotic superpotentials, and we explain how the various  physical and mathematical notions relate. In Section \ref{genproc} we describe how to use real bundle isomorphisms between the tangent and gauge bundles of heterotic compactifications to compute the holomorphic Chern-Simons invariant (\ref{phcs}). We also provide a concrete example of such a computation.  Section \ref{whyvanish} reviews an important theorem that explains why the Chern-Simons contribution vanishes in many cases. In Section \ref{fracsec} we discuss issues that arise when considering holomorphic Chern-Simons contributions to the superpotential in compactifications on quotient manifolds. In that section we also construct an explicit example of a heterotic compactification with a non-flat gauge bundle that gives rise to a fractional holomorphic Chern-Simons invariant. Finally, in Section \ref{conc} we briefly conclude and discuss possible future directions of research. The appendices contain several technical results that are necessary for our discussion.


\section{Basics of Chern-Simons terms}\label{sec:basics}

The reader may well be used to defining Chern-Simons invariants in the form discussed in the introduction. In many physical applications such a definition suffices. However, in the case of heterotic compactifications, the non-trivial topological structure of the compactification means that more care is required. In what follows we will compare the definitions of such invariants as they appear in the physics and mathematics literature, and we will describe why caution is required.

\subsection{Heterotic Chern-Simons terms}
The Chern-Simons term $\CS_{\rm phys}(A,\omega)$ which appears in heterotic theories and forms part of the heterotic superpotential has already been defined in (\ref{phcs}). How does this Chern-Simons term behave under gauge transformations
\begin{equation}
 A\mapsto hAh^{-1}+hdh^{-1}\;,\qquad \omega\mapsto g\omega g+gdg^{-1}\; ,
\end{equation}
of the gauge connection $A$ and the spin connection $\omega$? A short calculation reveals the Chern-Simons forms change as
\begin{equation} \label{phystrans}
\omega_3(A) \mapsto \omega_3(A)+ \tr \left[ d(A dh^{-1}h)-\frac{1}{3}(hdh^{-1})^3 \right] \; ,
\end{equation}
and similarly for $\omega_3(\omega)$, but with $h$ replaced by $g$. It is easy to see that the integrand of the Chern-Simons term $\CS_{\rm phys}(A,\omega)$ in (\ref{phcs}) is not invariant under these transformations. This is problematic since contributions to the integral~(\ref{phcs}) from two patches can differ on their overlap. In other words, the integral is not well-defined globally as it depends on the choice of partition of unity that is used in its definition. 

In the context of supergravity this problem is addressed by assigning a gauge transformation to the two-form field $B$ which cancels the variation~(\ref{phystrans}) so that the field strength $H$ is gauge invariant. This means that the sum of the first and second integral on the right-hand side of (\ref{GVW}) is gauge invariant, so the superpotential is well-defined as it should be. However, for the purpose of investigating the effect of the Chern-Simons contribution this state of affairs is quite inconvenient. It is desirable to have a well-defined version of the Chern-Simons term  and a way to express the superpotential~(\ref{GVW}) in terms of this object. We will define this mathematical version of the Chern-Simons term - the holomorphic Chern-Simons invariant - in the next sub-section and subsequently describe its relationship to the physics Chern-Simons term.

Before we do so, it is important to note that compactifications of heterotic string theory also contains another type of Chern-Simons integral defined over real three-manifolds. Heterotic flux quantization \cite{Rohm:1985jv,Lukas:1997rb} can be stated as the condition
\begin{equation} \label{fq}
\frac{1}{\alpha'} \int_{\cal C} H -  \int_{\cal C} \left(\omega_3(\omega) - \omega_3(A) \right)\in\mathbb{Z}
\end{equation}
for any integral three-cycle ${\cal C}\subset X$. Note that this condition involves an integral over a three-cycle in the Calabi-Yau space, in contrast  to the Chern-Simons term~(\ref{phcs}) which requires integration over the entire manifold. The integrand in~(\ref{fq}) is not gauge variant under the transformation~(\ref{phystrans}), and is, hence, ill-defined, much as its six-dimensional counterpart~(\ref{phcs}). To formulate flux quantization properly, we will introduce the ordinary Chern Simons invariant and subsequently explain how it enters the physical condition.

\subsection{Chern-Simons invariants} \label{sec:csi}
We begin by formulating the holomorphic Chern-Simons invariant, the object which will provide us with a well-defined version of the Chern-Simons term~(\ref{phcs}) which appears in the heterotic superpotential. Useful discussions of this and related topics, intended for an audience of physicists, can be found here \cite{Freed:1986zx,Dijkgraaf:1989pz,Freed:2008jq}. The set-up requires two connections\footnote{For brevity of exposition, we will sometimes conflate connections and the local gauge fields they give rise to in discussions where this should not cause confusion.}, $A$ and $A_0$, on the {\it same} vector bundle $V$ over a base Calabi-Yau manifold $X$. The connection $A$ will be seen as the argument of the Chern-Simons invariant and $A_0$ is called a ``reference connection". Then, with the adjoint valued one form $a=A-A_0$, the definition of the holomorphic Chern-Simons invariant is as follows \cite{Thomas:1998uj}.
\begin{eqnarray}\label{mhcs}
\CS_{A_0}(A) = \int_X \tr \left( (\overline{\partial}_{A_0} a\wedge a )+ \frac{2}{3} a \wedge a \wedge a + 2 a \wedge F_0 \right) \wedge \Omega
\end{eqnarray}
Here, $F_0$ is the field strength associated to the connection $A_0$ and we define the covariant derivative $d_{A_0}  a = d a + A_0 \wedge a+ a\wedge A_0$. Note that, naively, this is quite different from its supposed counterpart~(\ref{phcs}) in the heterotic theory which is defined in terms of two connections {\it seemingly on different bundles}, the gauge bundle and the tangent bundle. We will review the relationship between the mathematical and physics picture in the next sub-section. For now, we note that the holomorphic Chern-Simons invariant can be written in terms of the Chern-Simons forms $\omega_3(A)$ and $\omega_3(A_0)$, defined as in (\ref{Hdef}), as
\begin{eqnarray} \label{mhcs2}
\CS_{A_0}(A) = \int_X \tr \left( \omega_3 (A) - \omega_3 (A_0) - d (A \wedge A_0) \right) \wedge \Omega  \;.
\end{eqnarray}

What happens to the holomorphic Chern-Simons invariant under simultaneous gauge transformations of $A$ and $A_0$,
\begin{equation}\label{gtboth}
A \mapsto h A h^{-1} + h dh^{-1}\quad \textnormal{and}\quad A_0 \mapsto h A_0 h^{-1}+h dh^{-1}\; ,
\end{equation}
with the {\em same} gauge parameter $h$? Evidently, under such a gauge transformation all of the quantities appearing in (\ref{mhcs}) transform in a covariant manner and, as a result, the holomorphic Chern-Simons invariant (\ref{mhcs}) is thus manifestly invariant. Note, the transformation (\ref{gtboth}) is different to just performing a gauge transformation on $A$ while keeping $A_0$ fixed, a perhaps more familiar case which we will discuss shortly. The fact that the integrand in (\ref{mhcs}) is invariant under (\ref{gtboth}) means the integral is well-defined. More specifically, since the values of $A$ and $A_0$ on overlaps are related by gauge transformations and diffeomorphisms, the value of the integrand is well-defined everywhere. In practice, the integral can then be evaluated by combining the contributions from different patches with any suitable partition of unity.\\[2mm]
The holomorphic Chern-Simons invariant satisfies a number of properties which can be directly derived from its definition~(\ref{mhcs}) or the equivalent expression~(\ref{mhcs2}) and which will be useful for our subsequent discussion. First, for three connections $A$, $B$ and $C$ on the same bundle we have
\begin{equation}\label{hcsprop}
 \CS_B(A)=-\CS_A(B)\;,\qquad \CS_C(A)=\CS_B(A)-\CS_B(C)\; .
\end{equation} 
Further, the holomorphic Chern-Simons invariant  is unchanged under holomorphic deformations of the gauge connection. Consider an infinitesimal deformation, $\delta a=A-A_0$ of the connection $A_0$ to a connection $A$, so that
\begin{equation} \label{oneform}
 \textnormal{CS}_{A_0}(A)= 2\int_X \tr \left(  \delta a \wedge F_0\right) \wedge \Omega \;.
\end{equation}
Clearly, this expression vanishes if the connection $A_0$ is holomorphic. Thus holomorphic connections are extrema of the Chern-Simons functional. Any, even finite, deformation $A$ of $A_0$ which preserves the condition $F_{(0,2)}=0$ everywhere along a path in connection space from $A_0$ to $A$ will therefore lead to a vanishing Chern-Simons invariant, $\textnormal{CS}_{A_0}(A)=0$.

Computing the Chern-Simons invariant in heterotic models will often involve breaking up the deformation from the reference connection $A_0$ to a connection $A$ into several parts. Specifically, consider the sequence of deformations
\begin{equation}
 A_0\rightarrow A_1\rightarrow\cdots\rightarrow A_n=A\; .
\end{equation} 
Then, (\ref{hcsprop}) implies that the holomorphic Chern-Simons invariant is additive, in the sense
\begin{equation}\label{CSsum}
 \CS_{A_0}(A)=\sum_{k=1}^n\CS_{A_{k-1}}(A_k)\; .
\end{equation} 
Note that any of the partial deformations $A_{k-1}\rightarrow A_k$ which is holomorphic satisfies $\CS_{A_{k-1}}(A_k)=0$ and, hence, does not contribute the above sum.\\[2mm]
The other type of Chern-Simons invariant we need to introduce is the ordinary Chern-Simons invariant, defined by
\begin{eqnarray} \label{ocsdef}
\textnormal{OCS}_{A_0}(A,{\cal C}) = \int_{{\cal C}} \tr \left( (d_{A_0} a \wedge a )+ \frac{2}{3} a \wedge a \wedge a + 2 a \wedge F_0 \right)\; .
\end{eqnarray}
Here ${\cal C}$ is some three manifold, which will be a three-cycle within the Calabi-Yau three-fold $X$ in our application, and $A$ and $A_0$ are connections on a bundle $V$ over ${\cal C}$. Following exactly the same logic as  for the holomorphic Chern-Simons invariant, this object is invariant under the transformations (\ref{gtboth}) acting on both $A$ and $A_0$ simultaneously and with the same gauge parameter. This means the integral in (\ref{ocsdef}) is well defined.\\[2mm]
It will be crucial in Section \ref{fracsec} to understand how the two Chern-Simons invariants defined above behave under a different type of gauge transformation, namely one where $A$ changes by a large gauge transformation, while the reference connection $A_0$ is kept fixed.  To this end we reproduce here the standard construction from the literature addressing this issue \cite{donaldsonbook}.

Consider constructing a bundle $\mathbb{V}$ on ${\cal C} \times S^1$, where the circle is described by the interval $[0,1]$ with ends identified, by using a large gauge transformation $g$ to glue ${\cal V}|_{{\cal C} \times \{ 0\}}$ to ${\cal V}|_{{\cal C} \times \{ 1\}}$. We also construct another bundle $\mathbb{V}_0$ on the same four manifold by using the identity group element, rather than $g$, in the identification. We take $\mathbb{A}$ to be any connection on $\mathbb{V}$ that restricts to be $A$ on ${\cal C} \times \{0\}$ and therefore $g(A)$ on ${\cal C} \times \{1\}$.  We take $\mathbb{A}_0$ to be any connection on $\mathbb{V}_0$ that restricts to $A_0$ on both ${\cal C} \times \{0\}$ and ${\cal C} \times \{1\}$. Then, by Stokes' theorem, we have the following.
\begin{eqnarray} \label{lgtb}
\textnormal{OCS}_{A_0}(g(A),{\cal C}) - \textnormal{OCS}_{A_0}(A,{\cal C}) = \int_{{\cal C} \times S^1}\tr \left( \mathbb{F} \wedge \mathbb{F} \right) - \tr \left( \mathbb{F}_0 \wedge \mathbb{F}_0\right)
\end{eqnarray}
This is the usual statement that Chern-Simons invariants change by an integer under large gauge transformations. Note that we are viewing the situation in two different manners here. To use Stokes' theorem we are dropping the gluing with $g$ to simply have a line interval with boundary in order to obtain (\ref{lgtb}). On this space, $A$ and $A_0$ are connections on the same bundle so that the Chern-Simons invariant is well defined and we can use Stokes' theorem. Then, to claim that the right hand side of (\ref{lgtb}) is an integer we are viewing the situation as the glued geometry described above. We can then properly define the two topological invariants that appear independently as being associated to two different bundles and, since they are integrated over a closed manifold, we can see they are proportional to integers.

Thus, the ordinary Chern-Simons invariant (\ref{ocsdef}) changes by integers under large gauge transformations of their argument, with the reference connection fixed. It should also be pointed out that the fact that the reference connection is not transformed to obtain this behaviour is implicit in common applications of Chern-Simons invariants where $A_0$ is taken to vanish. If this were not the case one would obtain a non-vanishing reference connection after the gauge transformation.

A very similar construction to the one given above can be used to determine how the holomorphic Chern-Simons invariant behaves under a large gauge transformations of its argument with the reference connection fixed \cite{thomasthesis}. The result is that  $CS_{A_0}(A)$  changes by a period of the holomorphic three-form $\Omega$.\\[2mm]
The final question we would like to review in this sub-section is how the two types of Chern-Simons invariants, (\ref{mhcs}) and (\ref{ocsdef}), are linked. This is a little more difficult to see than in the case of flat bundles where the Chern-Simons invariants are closed \cite{Conrad:2000tk,Gukov:2003cy,Apruzzi:2014dza}, but is still straightforward.

We begin with a remark about the structure of the holomorphic Chern-Simons invariant. From (\ref{mhcs}), this invariant contains the form $ {\rm tr}(\overline{\partial}_{A_0} a\wedge a + \frac{2}{3} a \wedge a \wedge a + 2 a \wedge F_0)$ which is not closed. However, by the Hodge decomposition any form can be written as a sum of a closed and a co-exact form. Luckily, in the expression (\ref{mhcs}) for the holomorphic Chern-Simons invariant the co-exact piece does not contribute to the integral. To see this, write this co-exact piece as $d^{\dagger} \beta_4$ and work out its contribution to the Chern-Simons invariant which is given by
\begin{equation}
\int d^{\dagger} \beta_4 \wedge \Omega= -i \int d^{\dagger} \beta_4 \wedge * \Omega = -i \left< d^{\dagger} \beta_4 , \overline{\Omega} \right> = -i \left< \beta_4,d \overline{\Omega} \right> =0\; .
\end{equation}
Given this, we can treat the holomorphic Chern-Simons invariant as an integral over a wedge product of two closed forms. In fact this is, in part, why the Chern-Simons term is so hard to compute. The information we have easy access to - the relation of $\omega_3(A)$ to $\tr(F\wedge F)$ - drops out of the integral defining the invariant.

This being the case, let us take the usual symplectic basis of the third cohomology of $X$, $(\alpha_i,\beta^i)$, and the associated dual basis of three-cycles $({\cal A}^i,{\cal B}_i)$. These quantities obey the following standard special geometry relations
\begin{equation}\begin{array}{rllllll}
\int_X \alpha_i \wedge \beta^j &=& \int_{{\cal A}^j} \alpha_i = \delta^j_i&\qquad\quad& \int_X \alpha_i \wedge \alpha_j &=&0 \\[2mm]
\int_X \beta^j \wedge \alpha_i &=& \int_{{\cal B}_i} \beta^j = - \delta^j_i&\qquad\quad& \int_X \beta^i \wedge \beta^ j &=&0
\end{array}\; .
\end{equation}
In terms of the cohomology basis, we can expand the holomorphic three-form $\Omega$ as
\begin{eqnarray}
 \Omega = {\cal Z}^i \alpha_i - {\cal G}_i \beta^i\; ,
\end{eqnarray}
where ${\cal Z}^i$ are the usual coordinates on the complex structure moduli space and the ${\cal G}_j$ are the derivatives of the pre-potential with respect to these variables. Given this set-up, the holomorphic Chern-Simons invariant can now be expressed in terms of ordinary Chern-Simons invariants associated to the basis three-cycles $({\cal A}^i,{\cal B}_i)$. A short calculation shows that
\begin{equation} \label{breakdown}
\textnormal{CS}_{A_0}(A) = b_i {\cal Z}^i - a^i {\cal G}_i\;,\qquad a^i=\textnormal{OCS}_{A_0}(A,{\cal A}^i)\;,\qquad b_i=\textnormal{OCS}_{A_0}(A,{\cal B}_i)\; .
\end{equation}
Hence, the ordinary Chern-Simons invariants, carried out over a basis of three-cycles, determine the holomorphic Chern-Simons invariant. Note that this result is consistent with the above discussion of how these objects behave under a large gauge transformation. Under such large gauge transformations, the ordinary Chern-Simons invariants and, hence, the numbers $a_i$ and $b^i$, change by integers. Equation (\ref{breakdown}) then implies that the holomorphic Chern-Simons invariant changes by a period, in agreement with the earlier discussion.

We can now be more precise about what we mean by a ``fractional holomorphic Chern-Simons invariant". This terminology indicates a holomorphic Chern-Simons invariant for which at least one of the numbers $(a^i,b_i)$ in (\ref{breakdown}) is not an integer. An analogous definition has been used in related work studying flat bundles \cite{Conrad:2000tk,Gukov:2003cy,Apruzzi:2014dza}. 

\subsection{Chern-Simons invariants in heterotic theories}
We now have the tools to examine how the Chern-Simons invariants introduced in the previous sub-section relate to the Chern-Simons terms which appear in heterotic theories and how we can use the former to calculate the latter. This correspondence will form the basis of our subsequent calculations.\\[2mm]
The first difference to resolve is the apparent discrepancy in the set-up of vector bundles. While the physical Chern-Simons term~(\ref{phcs}) depends on connections with independent gauge transformations on apparently different bundles, the tangent bundle $TX$ and the gauge bundle $V$ on $X$, the holomorphic Chern-Simons term~(\ref{mhcs}) depends on a connection and a reference connection, both defined on the {\em same} bundle. What comes to the rescue is the fact that two $E_8$ bundles are the same {\it as real bundles}\footnote{Note that we use the term ``real bundle" here to refer to the underlying smooth structure (i.e. real bundle as opposed to complex bundle), not to refer to a real structure on a holomorphic bundle (e.g. special orthogonal or symplectic structures).} if and only if their second Chern characters, as elements of $H^4(X,\mathbb{Z})$, match \cite{Witten:1985bt,Wen:1985qj}. In the case where no five-branes are present and we only have a bundle in one $E_8$ factor of the heterotic gauge group the second Chern characters for the tangent bundle and the gauge bundle must be equal~\footnote{Note that, in order for heterotic theory to be well defined more generally, there must be a generalization of the consistent mathematical definition of the Chern-Simons invariant to include more complicated cases, such as those involving contributions to the Bianchi Identity from M5 branes. Such generalizations would certainly be interesting to pursue, both from the perspective of physics and mathematics, but are beyond the scope of the current work.}. This follows from the integrability condition on the heterotic Bianchi identity (\ref{gbi}) for the Chern-Characters with real coefficients,
and from global worldsheet anomaly considerations for the extension to include torsion \cite{Witten:1985mj,Witten:1999eg}. Therefore the tangent bundle and gauge bundle are the same {\it as real bundles} in such a situation. Note this does not mean that they are the same as holomorphic objects. Indeed this could not be possible in the case, for example, where the third Chern class of the gauge bundle differs from that of the tangent bundle. 

Given this discussion, the physical and mathematical versions of the Chern-Simons invariant start to look somewhat similar. Only two differences remain. The first is that in the holomorphic Chern-Simons invariant~(\ref{mhcs}), both $A$ and $A_0$ are defined relative to the same trivialization of the bundle. This is in distinction to (\ref{phcs}) where they would be written with respect to two different trivializations adapted to the holomorphic structure on the gauge and tangent bundles respectively. Second, (\ref{mhcs}) contains one additional term relative to (\ref{phcs}) which, despite initial appearances, cannot be integrated by parts to obtain zero.

To address these differences, we recall that the $H$-part of the superpotential
\begin{equation}
 W_H=\int_X H\wedge\Omega=\int_XH_0\wedge\Omega+\alpha' \CS_{\rm phys}(A,\omega)
\end{equation}
is actually gauge invariant due to the gauge invariance of $H$. The invariance is achieved by cancelling the non-vanishing variation of the Chern-Simons term against the variation of $H_0$ which can be written as $dB$ locally. Hence, we can choose a gauge where $A$ and $\omega$, which we can now think of as connections on the same bundle, are described relative to the same trivialization. Further, in order to remove the additional term which appears in (\ref{mhcs}) relative to (\ref{phcs}) we can use a gauge where at least one of the connections $A$ and $A_0$ has a vanishing $(0,1)$ component. This is always possible because the connections of physical interest can be written as Chern-connections in appropriate trivializations. Having fixed a gauge in this manner, we can express $W_H$ in terms of the holomorphic Chern-Simons invariant as
\begin{equation}
 W_H=\int_X H_0\wedge\Omega + \alpha' \CS_\omega (A)\;  .\label{WH}
 \end{equation}
We already know from (\ref{breakdown}) that this superpotential can be expressed in term of ordinary Chern-Simons invariants, associated to the symplectic basis $({\cal A}^i,{\cal B}_i)$ of three-cycles. More precisely, the two terms in (\ref{WH}) can be written as
\begin{equation}\label{WH2}
 \frac{1}{\alpha'} \int_X H_0\wedge\Omega =m_i{\cal Z}^i-n^i{\cal G}_i\;,\qquad  \CS_\omega (A)=b_i{\cal Z}^i-a^i{\cal G}_i \; ,
\end{equation}
where
\begin{equation} 
\begin{array}{lllllll}
  a^i&=&\textnormal{OCS}_\omega(A,{\cal A}^i)&\qquad& b_i&=&\textnormal{OCS}_\omega(A,{\cal B}_i)\\[2mm]
  n^i&=&\frac{1}{\alpha'} \int_{{\cal A}^i}H-\textnormal{OCS}_\omega(A,{\cal A}^i)&\qquad&m_i&=&\frac{1}{\alpha'}\int_{{\cal B}_i}H-\textnormal{OCS}_\omega(A,{\cal B}_i)
\end{array} \; .
\end{equation}  
Consequently, the full superpotential $W_H$ is given by
\begin{equation}
 \frac{1}{\alpha'} W_H=(m_i+b_i){\cal Z}^i-(n^i+a^i){\cal G}_i\; . \label{WHfinal}
\end{equation}
The flux quantization condition~(\ref{fq}), properly expressed in terms of ordinary Chern-Simons invariants, takes the form
\begin{equation}
 \frac{1}{\alpha'} \int_{{\cal C}}H-\textnormal{OCS}_\omega(A,{\cal C})\in\mathbb{Z}
\end{equation} 
and it shows that the quantities $n^i$ and $m_i$ in (\ref{WHfinal}) are, in fact, integers. In other words, the integers $n^i$ and $m_i$ describe the harmonic flux in $H_0$ while the potentially fractional quantities $a^i$ and $b_i$ describe the holomorphic Chern-Simons invariant. It is the latter, which are the main subject of this paper.

As is clear from the above discussion, a key ingredient in computing these quantities in actual heterotic models is knowing the isomorphism between the tangent bundle and the vector bundle explicitly. This is needed to write the connections on the tangent and gauge bundles relative to the same trivialization. Given that this isomorphism is typically not holomorphic it is not easy to find and we will describe in Section \ref{genproc} and Appendix \ref{realbunmorpheg} how this can be done in certain cases.


\section{Calculating Chern-Simons invariants} \label{genproc}

\subsection{General approach} \label{sec:genapp}
After the general discussion of the last section we return to the central goal of this paper. We want to compute the holomorphic Chern-Simons invariant (\ref{mhcs}) for specific connections over Calabi-Yau three-folds $X$ that appear in heterotic compactifications. In particular, we are interested in the case where $A_0=\omega$ is the spin connection on $TX$ and $A$ the gauge connection on a bundle $V\rightarrow X$ solving the Hermitian Yang-Mills equations
\begin{equation} \label{hym}
F_{\overline{a}\overline{b}} =0\;,\qquad g^{a \overline{b}} F_{a \overline{b}} =0\; .
\end{equation}
We will choose to write these connections in ``math gauge" as Chern connections. Note that different gauge transformations would be needed on $A$ and $A_0$ in order to write them in the gauge, more prevalent in the physics literature, where these fields are real. Given the transformation properties discussed in Section \ref{sec:basics}, this means that the result we will obtain will generically change by an integer if we chose to do this. Obviously, such an integer shift cannot change whether or not a Chern-Simons invariant is fractional, which is a main point of interest here.

Typically, the holomorphic structure of the gauge and tangent bundles in a heterotic compactification are different. The Chern connection solving (\ref{hym}) and the spin connection would be written in terms of different local trivializations respecting these structures. Nevertheless, we could compute the holomorphics Chern-Simons invariant (\ref{mhcs}) if we had the requisite real bundle isomorphisms between the two bundles. Let us discuss how such a computation would proceed.\\[2mm]
Let us phrase this discussion more generally, in terms of two bundles $V\rightarrow X$ and $V'\rightarrow X$ over a Calabi-Yau three-fold $X$ and a (possibly non-holomorphic) bundle isomorphism $f:V'\rightarrow V$. One might imagine taking $V'=TX$, for example, given the above discussion. However, we wish to keep our notation more general because, as we will see, in practice this might be required for the computation. We assume that we have connections $\nabla_0$ and $\nabla'$ on $V$ and $V'$, respectively, as well as local frames $s_a$ and $s_a'$ associated to some given open set in the base. Then, relative to these local frames, the gauge fields $A_0$ and $A'$ associated to $\nabla_0$ and $\nabla'$ are obtained from
\begin{equation}
 \nabla_0 s_a={A^b}_{0a} s_b\;,\qquad \nabla' s_a'={A^{\prime b}}_a s'_b\; .
\end{equation} 
We can use the bundle morphism $f$ to ``transport" the connection $\nabla'$ on $V'$ to a connection on $V$, which we will denote by $\nabla$.  This connection is defined by
\begin{equation}
 \nabla(s):=f\circ\nabla'(f^{-1}\circ s)\; ,
\end{equation} 
where $s$ is a section of $V$. The bundle morphism $f$ can also be used to map the frame $s_a'$ of $V'$ to a frame $\tilde{s}_a:=f\circ s_a'$ of $V$. Now we have two frames, $s_a$ and $\tilde{s}_a$ on $V$ and thus there is a gauge transformation
\begin{equation} \label{frameact}
 s_a={P^b}_a\tilde{s}_b
\end{equation}
relating them. We can work out the gauge field which corresponds to $\nabla$ relative to the frame $\tilde{s}_a$ and the frame $s_a$. The result is
\begin{equation} \label{Adef}
 \nabla(\tilde{s}_a)={A^{\prime b}}_a\tilde{s}_b\;,\qquad \nabla(s_a)={A^b}_as_b\quad\textnormal{where}\quad A=P^{-1}A'P+P^{-1}dP\; .
\end{equation} 
In other words, relative to the frame $\tilde{s}_a$, obtained by transporting the frame on $V'$ to $V$, the gauge field remains unchanged, that is, it is given by $A'$  for both the frames $s_a'$ on $V'$ and $\tilde{s}_a$ on $V$. For the frame $s_a$ on $V$, on the other hand, the gauge field is obtained from $A'$ by the above gauge transformation.\\[2mm]
Now that we have phrased matters in terms of two connections on the same bundle, we can work out the holomorphic Chern-Simons term more explicitly.  Suppose that both initial connections $A'$ and $A_0$ are Chern connections. (But note that $A$, being obtained from $A'$ by a potentially non-holomorphic bundle morphism, does not need to be a Chern connection.) Then, the term $d(A \wedge A_0) \wedge \Omega$ in (\ref{mhcs2}) vanishes simply by index structure arguments and a quick calculation shows the remaining terms satisfy
\begin{equation}
 \omega(A)-\omega(A_0)=\omega(A')-\omega(A_0)+{\rm tr}\left(\theta\, dA'-A'\theta^2-\frac{1}{3}\theta^3\right)\;,\qquad \theta:=dP\, P^{-1}\; .
\end{equation} 
In particular, in our case where $A'$ and $A_0$ happen to be $(1,0)$ gauge fields, we have
\begin{equation} \label{chernconresult}
\left[ \omega(A)-\omega(A_0)\right]_{(0,3)}=-\frac{1}{3}{\rm tr}(\theta^3)_{(0,3)}\qquad \textnormal{so that} \qquad
\textnormal{CS}_{A_0}(A)=-\frac{1}{3}\int_X{\rm tr}(\theta^3)\wedge\Omega\; .
\end{equation}
Thus we see from (\ref{chernconresult}) that if the real isomorphism $f$ is known, so that $P$ can be obtained from its action on frames via (\ref{frameact}), then we can compute the Chern-Simons invariant associated to Chern connections $A$ and $A_0$, even if we do not know the explicit form of these connections themselves. Clearly this his helpful given the non-constructive nature of the Yau \cite{yau} and Donaldson-Uhlenbeck-Yau theorems \cite{duy1,duy2}.

It should be noted that, given the form of the result (\ref{chernconresult}), one might expect this Chern-Simons invariant to not be fractional since the integrand looks like the wedge product with $\Omega$ of the standard integrand giving a winding number. In fact, the holomorphic Chern-Simons invariant can be fractional as we will demonstrate in Section \ref{finaleg}. Nevertheless, even if this were to be the case an integral result here would still be important. A non-zero Chern-Simons invariant $\CS_{A_0}(A)$ of this type would destabilize the usual meta-stable vacuum in the absence of other effects. In addition, non-zero integer results can lead to fractional Chern-Simons invariants in quotients, as we will discuss in Section \ref{fracsec}. 

\subsection{Finding the isomorphism} \label{sec:iso}
It is clear from the proceeding discussion that the key quantity we need to compute is the real bundle isomorphism $f$ between $V'$ and $V$ and the associated gauge transformations $P$. How do we describe such an isomorphism practically? Since this is somewhat technical, a full description of this topic is relegated to Appendix \ref{realbunmorpheg}. Here we will content ourselves with a summary of the essential ideas, along with a simple illustrative example.\\[2mm]
By definition, vector bundles locally look like a direct product of an open set on the base manifold and the fiber. In other words, we have local trivializations,
\begin{eqnarray} \label{localtriv}
\phi_{\alpha}: \pi^{-1}(U_{\alpha}) \to W_{\alpha} \times F \;.
\end{eqnarray}
Here $\pi$ is the projection map of the bundle and $W_{\alpha}$ is an open subset of $\mathbb{C}^{{\rm dim}(X)}$ and $F\cong\mathbb{C}^{{\rm rk}(V)}$ is the typical fiber. These local trivializations are glued together by transition functions $\phi_{\alpha \beta} := \phi_{\alpha} \circ \phi^{-1}_{\beta}: W_{\beta}\times F \to W_{\alpha} \times F$ to construct the bundle globally. The transition functions act trivially on the base and as linear maps $T_{\alpha \beta}$ on the fiber $F$.

In terms of these local trivializations, our real bundle isomorphism $f$ is described by a collection of maps 
\begin{eqnarray}.
f_{\alpha} : W_{\alpha} \times F \to W_{\alpha} \times F\; .
\end{eqnarray}
Because we want the isomorphism to act fiber-wise, preserving points on the base, these maps take the form $f_{\alpha}(z,v) = (z, P_{\alpha} v)$ where the $z$ are coordinates on the open set $W_{\alpha}$ in the base and the $v$ are coordinates in the fiber $F$. In short, the real bundle isomorphism $f$ can be described by a collection of matrices $P_{\alpha}$ which encodes, for each patch, how the fibers of $V'$ are mapped to those of $V$. In fact, these are precisely the matrices appearing in (\ref{frameact}).

The matrices $P_{\alpha}$ must satisfy several consistency conditions. The first is that, if they are to map  $V'$ to $V$, then they must correctly map the transition functions of the first bundle into those of the second. That is, they must obey the intertwining conditions
\begin{eqnarray} \label{transsimp}
 T_{\alpha\beta}= P_{\alpha}^{-1} {T'}_{\alpha\beta}P_\beta
\end{eqnarray}
for all patches $\alpha,\beta$. All matrices $P_{\alpha}$ must also be invertible (to define an isomorphism rather than just a morphism) and they must be non-singular (to be well defined). We describe all of these conditions in detail in Appendix \ref{realbunmorpheg}. The non-holomorphic nature of the bundle morphisms we will utilize manifests itself in the fact that the matrices $P_{\alpha}=P_{\alpha}(z,\overline{z})$ are, in general, not holomorphic functions of the base coordinates.\\[2mm]
Reverting the logic of the discussion, we can say that any collection of matrices $P_\alpha$, all invertible and non-singular, which satisfy the intertwining conditions~(\ref{transsimp})
{\it define} a bundle morphism $f$. Thus, in order to compute the holomorphic Chern-Simons invariant $CS_{A_0}(A)$ via (\ref{chernconresult}) we need to obtain such a set of matrices $P_\alpha$.\\[2mm]
Let us illustrate this discussion with a concrete example on the simple base manifold $\mathbb{P}^1$. It is known that line bundle sums on $\mathbb{P}^1$ are classified by their total Chern character. In particular, this means that the line bundle sums
\begin{equation}\label{VV}
 V'={\cal O}_{\mathbb{P}^1}(-1) \oplus {\cal O}_{\mathbb{P}^1}(1)\quad\mbox{and}\quad  V={\cal O}_{\mathbb{P}^1} \oplus {\cal O}_{\mathbb{P}^1}
 \end{equation}
which both have vanishing first Chern class are real isomorphic. How do we write down such an isomorphism in the form we have been discussing? The standard open cover of $\mathbb{P}^1$ has two patches, which we label $U_0$ and $U_1$ with affine coordinates $z$ and $w$, respectively. Relative to those patches, the transition functions for the bundles $V'$ and $V$ in (\ref{VV}) are
\begin{equation}
 T'_{10}=\textnormal{diag}(z,z^{-1})\;,\qquad T_{10}=\textnormal{diag}(1,1)\; .
\end{equation} 
Then, two matrices $P_{\alpha}$ which satisfy the intertwining conditions~(\ref{transsimp}) with these transition functions  can simply be written as 
\begin{equation} \label{simpeg}
  P_0(z,\bar{z})=\left(\begin{array}{cc}1&\frac{\bar{z}}{1+|z|^{2}}\\-z&\frac{1}{1+|z|^{2}}\end{array}\right)\;,\qquad P_1(w,\bar{w})=\left(\begin{array}{cc}w&\frac{1}{1+|w|^{2}}\\-1&\frac{\bar{w}}{1+|w|^{2}}\end{array}\right)\; .
\end{equation}
These are clearly non-singular and invertible in their respective patches. Note that the matrices in (\ref{simpeg}) depend upon both the complex coordinates and their conjugates. This had to be the case since the bundles~(\ref{VV}) are not isomorphic as holomorphic objects.\\[2mm]
The above construction can be generalized to relate any two rank two line bundle sums on $\mathbb{P}^1$ with the same first Chern class. The resulting bundle morphisms have the following structure
\begin{eqnarray}
 f^{(q,p)}\sim\left(P^{(q,p)}_\alpha\right)&:&{\cal O}_{\mathbb{P}^1}(a-p)\oplus{\cal O}_{\mathbb{P}^1}(a+p)\stackrel{\simeq}{\longrightarrow} 
 {\cal O}_{\mathbb{P}^1}(a-q)\oplus{\cal O}_{\mathbb{P}^1}(a+q) \label{fmorpheven}\\
 g^{(q,p)}\sim\left(Q^{(q,p)}_\alpha\right)&:& {\cal O}_{\mathbb{P}^1}(a-p)\oplus{\cal O}_{\mathbb{P}^1}(a+p+1)\stackrel{\simeq}{\longrightarrow} 
 {\cal O}_{\mathbb{P}^1}(a-q)\oplus{\cal O}_{\mathbb{P}^1}(a+q+1)\; ,\label{fmorphodd}
\end{eqnarray} 
for even and odd first Chern classes, respectively. Their explicit form is a generalization of (\ref{simpeg}) and is provided in Appendix \ref{realbunmorpheg}, where we explain this construction in more detail. Of course, these results cannot be applied to our problem directly but, as we will see, they can be used to construct bundle isomorphisms on Calabi-Yau manifolds which are defined in ambient spaces that involve $\mathbb{P}^1$ factors.\\[2mm]

\subsection{An explicit example} \label{mr0}

In this section we will work on the tetra-quadric Calabi-Yau three-fold, defined as the zero-locus of a polynomial of multi-degree $(2,2,2,2)$ in the ambient space $(\mathbb{P}^1)^4$, and represented by the configuration matrix
\begin{eqnarray} \label{X1}
X\in \left[ \begin{array}{c|c} \mathbb{P}^1 & 2 \\ \mathbb{P}^1 & 2 \\ \mathbb{P}^1 & 2 \\ \mathbb{P}^1 & 2 \end{array}\right] \; .
\end{eqnarray}
An appealing feature of this example is the presence of the $\mathbb{P}^1$ factors which, as we will see, allows us to transfer the results for real bundle equivalence on $\mathbb{P}^1$ to the tetra-quadric.\\[2mm]
Before we construct the relevant bundles on this manifold, we introduce the main building blocks
\begin{equation}\begin{array}{lllllllll}
 B&=&{\cal O}_X(1,0,-1,0)&\oplus&{\cal O}_X(1,1,0,0)&\oplus&{\cal O}_X(-1,1,0,0)&\oplus&{\cal O}_X(1,0,1,0)\\
\tilde{B}&=&{\cal O}_X(1,0,0,0)&\oplus&{\cal O}_X(1,0,0,0)&\oplus&{\cal O}_X(0,1,0,0)&\oplus&{\cal O}_X(0,1,0,0)\\
R&=&{\cal O}_X(0,0,1,0)&\oplus&{\cal O}_X(0,0,1,0)&\oplus&{\cal O}_X(0,0,0,1)&\oplus&{\cal O}_X(0,0,0,1)\\
C&=&{\cal O}_X(2,2,2,2)
\end{array} 
\end{equation} 
which underly our construction. Given these line bundle sums, we define the monad bundle $V$ on $X$ by
\begin{equation}\label{Vdef}
 0\longrightarrow V\longrightarrow B\oplus R\longrightarrow C\longrightarrow 0\; .
\end{equation} 
The Chern connection on $V$ is denoted by $A$ and our goal is to compute the holomorphic Chern-Simons invariant $\textnormal{CS}_\omega(A)$, relative the the spin connection $\omega$ on $TX$. To do this, we first find a way to deform the spin connection $\omega$ to the connection $A$.

Or first step is to introduce a monad representation of the tangent bundle
\begin{equation}
 0\longrightarrow V_0\longrightarrow \tilde{B}\oplus R\stackrel{\mu_0}{\longrightarrow} C\longrightarrow 0\; . \label{V0seq}
\end{equation} 
Indeed, for a suitable choice of the monad map $\mu_0$ we have $V_0\cong TX \oplus {\cal O}_X^{\oplus 4}$. We denote the Chern connection on $V_0$ by $A_0$. However, for different choices of the monad map the sequence~(\ref{V0seq}) describes holomorphic deformations away from the tangent bundle. In particular, we can choose $\mu_0$ such that the four line bundles in $\tilde{B}$ split off as a direct sum. This choice leads to a bundle $V_1$, with Chern connection $A_1$, which can be written as
\begin{equation}
 V_1=\tilde{B}\oplus U\;,\qquad 0\longrightarrow U\longrightarrow R\longrightarrow C\longrightarrow 0\; .
\end{equation} 
The next step is crucial. We use real bundle morphisms on $\mathbb{P}^1$, applied to our ambient space and restricted to the Calabi-Yau manifold, to construct a real bundle morphism ${\cal F}$ between the line bundle bundle sums $\tilde{B}$ and $B$. We will explain the procedure in more detail below but for now we continue outlining the structure of the argument.

Thanks to this real bundle morphism, we can relate the above bundle $V_1$ to the bundle
\begin{equation}
 V_2=B\oplus U
\end{equation}
with Chern connection $A_2$.  Evidently, this bundle is a holomorphic deformation of our gauge bundle $V$ in (\ref{Vdef}).

To summarize, we have now related the tangent bundle $TX$ to our gauge bundle $V$ via a number of deformations which can be schematically written as
\begin{equation}
\begin{array}{cllllllll}
 TX\oplus{\cal O}_X^{\oplus 4}&\stackrel{\rm hol.}{\longrightarrow}&V_0&\stackrel{\rm hol.}{\longrightarrow}&V_1&\stackrel{\rm real}{\longrightarrow}&V_2&\stackrel{\rm hol.}{\longrightarrow}&V\\
 \omega&\longrightarrow&A_0&\longrightarrow&A_1&\longrightarrow&A_2&\longrightarrow&A\; .
\end{array} 
\end{equation}
From (\ref{CSsum}) the holomorphic Chern-Simons invariant $\CS_\omega(A)$ can be computed by summing the holomorphic Chern-Simons invariants of the four steps in the above sequence. However, three of these steps correspond to holomorphic deformations. It is easy to see that these are not only holomorphic deformations at the level of the bundles, but are also holomorphic deformations of the Chern-connections. Hence, from (\ref{oneform}), the Chern-Simons invariants associated to these three steps vanish. In conclusion, the only contribution arises from the real deformation in the above sequence, so that
\begin{equation}
 \CS_\omega(A)=\CS_{A_1}(A_2)\; . \label{CSex}
\end{equation} 
We know from the general discussion in Section~\ref{sec:genapp} that $\CS_{A_1}(A_2)$ can be worked out from the bundle isomorphism ${\cal F}:\tilde{B}\rightarrow B$ so our next task is to construct this isomorphism.\\[2mm]
To do this, we recall from the previous subsection (see (\ref{fmorpheven})) that we can find explicit real bundle isomorphisms $f^{(q,p)}$ which relate pairs of rank two line bundle sums on $\mathbb{P}^1$ with the same first Chern class. For our present example, we have four $\mathbb{P}^1$ ambient space factors, which we label by $i=1,2,3,4$, as well as four line bundles in $\tilde{B}$, which we label by $a=1,2,3,4$. The maps $f^{(q,p)}$ can be applied to any of the four $\mathbb{P}^1$ factors and to any two of the four line bundles, while leaving the other $\mathbb{P}^1$ factors and line bundle undisturbed. We denote the version of $f^{(q,p)}$ which acts on the $i^{\rm th}$ $\mathbb{P}^1$ factors and on the two line bundles $a$ and $b$ by $f^{(q,p)}_{ab,i}$. As is evident, this gives rise to a large number of possibilities and a corresponding web of real bundle isomorphisms for line bundle sums on $(\mathbb{P}^1)^4$ (and, by restriction, on the tetra-quadric). It would be interesting to explore this more systematically.\\[2mm]
For present purposes, this formalism can be used to construct a real bundle isomorphism between $\tilde{B}$ and $B$ by the following chain.
\begin{equation}
\begin{array}{lllllllll}
 \tilde{B}&=&\left(\begin{array}{rrrr}1&0&0&0\\1&0&0&0\\0&1&0&0\\0&1&0&0\end{array}\right)&\stackrel{f^{(1,0)}_{12,2}}{\longrightarrow}&
                \left(\begin{array}{rrrr}1&-1&0&0\\1&1&0&0\\0&1&0&0\\0&1&0&0\end{array}\right)&\stackrel{f^{(1,0)}_{34,1}}{\longrightarrow}&
                 \left(\begin{array}{rrrr}1&-1&0&0\\1&1&0&0\\-1&1&0&0\\1&1&0&0\end{array}\right)&&\\
      &&&\stackrel{f^{(0,1)}_{14,2}}{\longrightarrow}&\left(\begin{array}{rrrr}1&0&0&0\\1&1&0&0\\-1&1&0&0\\1&0&0&0\end{array}\right)&\stackrel{f^{(1,0)}_{14,3}}{\longrightarrow}&\left(\begin{array}{rrrr}1&0&-1&0\\1&1&0&0\\-1&1&0&0\\1&0&1&0\end{array}\right)&=&B
 \end{array}                
\end{equation} 
Here, for ease of notation, we have written the line bundle sums as matrices, with each row representing the multi-degree of one line bundle. Hence, the desired line bundle isomorphism ${\cal F}:\tilde{B}\rightarrow B$ can be written as
\begin{equation} \label{Fmorph}
{\cal F} = f_{14,3}^{(1,0)} \circ f_{14,2}^{(0,1)} \circ f_{34,1}^{(1,0)}\circ f_{12,2}^{(1,0)} \;,
\end{equation}
suitably restricted to the Calabi-Yau three-fold $X$. It is straightforward to promote ${\cal F}$ to a bundle isomorphism $V_1\rightarrow V_2$ by extending it trivially onto the common summand $U$.\\[2mm]
The direct computation of the holomorphic Chern-Simons invariant~(\ref{CSex}), based on the formula~(\ref{chernconresult}), is not hard at this stage but simply tedious. We first compute the matrices $P_{\alpha}$ which represent the local descriptions of the bundle isomorphism (\ref{Fmorph}). For the purpose of integration, it is sufficient to carry this out in the standard patch of $(\mathbb{P}^1)^4$ whose affine coordinates we denote by $z_i$, where $i=1,2,3,4$. Combining the individual pieces in (\ref{Fmorph}), given in (\ref{fmorph}), (\ref{mrP}) and (\ref{P01def}),  we find the following expression for the local version of ${\cal F}$.
\begin{eqnarray} \label{egP}
P=\left(
\begin{array}{cccc}
 \frac{\left(\bar{z}_2+1\right) \left(z_2 \bar{z}_3+1\right)}{\left(|z_2|^2+1\right) \left(|z_3 |^2+1\right)} & -\frac{\left(\bar{z}_2-1\right) \left(z_2 \bar{z}_3+1\right)}{2 \left(|z_2|^2+1\right) \left(|z_3 |^2+1\right)} & \frac{\left(z_1-1\right) \left(\bar{z}_2-\bar{z}_3\right)}{\left(|z_2|^2+1\right) \left(|z_3 |^2+1\right)} & \frac{\left(z_1+1\right) \left(\bar{z}_2-\bar{z}_3\right)}{2 \left(|z_2|^2+1\right) \left(|z_3 |^2+1\right)} \\
 z_2-1 & \frac{1}{2} \left(z_2+1\right) & 0 & 0 \\
 0 & 0 & \frac{\bar{z}_1+1}{|z_1 |^2+1} & \frac{1-\bar{z}_1}{2 |z_1|^2+2} \\
 -\frac{\left(z_2-z_3\right) \left(\bar{z}_2+1\right)}{\left(|z_2|^2+1\right)} & \frac{\left(z_2-z_3\right) \left(\bar{z}_2-1\right)}{2 \left(|z_2|^2+1\right)} & \frac{\left(z_1-1\right) \left(z_3 \bar{z}_2+1\right)}{\left(|z_2|^2+1\right)} & \frac{\left(z_1+1\right) \left(z_3 \bar{z}_2+1\right)}{2 \left(|z_2|^2+1\right)} \\
\end{array}
\right)
\end{eqnarray}
To restrict to the Calabi-Yau three-fold we must pick a defining relation for the configuration~(\ref{X1}) and then solve it on the given patch for one of the coordinates in terms of the others. For example, the coordinate $z_4$ on the last $\mathbb{P}^1$ factor in the ambient space is a natural choice given the dependencies appearing in (\ref{egP}). Given the defining equation is a quadric in $z_4$, this yields two disjoint loci describing parts of $X$ inside the open patch, on which we know the matrix $P$.

Once we have computed $P$ we can then carry out the integral in (\ref{chernconresult}) in order to evaluate the Chern-Simons invariant, using the standard expression for the holomorphic $(3,0)$-form on such manifolds \cite{Witten:1985xc,Strominger:1985it,Candelas:1987se,Candelas:1987kf}. The integral one obtains vanishes.\\[2mm]
We have thus completed the first non-trivial computation of a holomorphic Chern-Simons invariant contribution to the heterotic superpotential, for the non-flat bundle (\ref{Vdef}) over the Calabi-Yau manifold (\ref{X1}), albeit obtaining a vanishing result. We have repeated such a computation for a large number of different gauge bundles over different Calabi-Yau manifolds.  The integrands all exhibit similar structures, not very different from those which arise in period integrals~\cite{Candelas:1990rm,Berglund:1993ax}, but the resulting Chern-Simons invariant always vanishes. At this stage one is motivated to look for more general reasons as to why a vanishing result might be obtained in many cases, or why fractional Chern-Simons invariants do not appear. This might provide some insight into our results so far and also guidance on how to build non-flat bundles with fractional contributions to the superpotential. We discuss a relevant vanishing theorem in the next section, and an example with a fractional holomorphic Chern-Simons invariant in Section \ref{fracsec}.

\section{A vanishing theorem and its consequences} \label{whyvanish}

In this section we will consider the consequences of the following theorem due to R. Thomas \cite{Thomas:1998uj}.
\begin{theorem} \label{tt}
Suppose that the Calabi-Yau three-fold $X$ is a smooth effective anti-canonical divisor in a four-fold $Y$ defined by $s \in H^0(K_Y^{-1})$. If $E\to X$ is a bundle that extends to a bundle $\mathbb{E} \to \mathbb{Y}$, then for a $\overline{\partial}$-operator $A$ on $E$, let $\mathbb{A}$ be any $\overline{\partial}$-operator on $\mathbb{E}$ extending $A$. Then we have, modulo periods and for some choice of reference connection,
\begin{eqnarray}
\textnormal{CS}(A) =  \int_Y \tr\left( \mathbb{F}_{0,2} \wedge \mathbb{F}_{0,2} \right) \wedge s^{-1} \;.
\end{eqnarray}
\end{theorem}
In terms of the notation in this theorem, the Chern-Simons invariant $\CS_\omega(A)$ of the connection $A$ on the bundle $V$, relative to the spin connection $\omega$ on the tangent bundle, can be written as $\CS_\omega(A)=\CS(A)-\CS(\omega)$. Clearly, this theorem has important implications for the questions being addressed in this paper. For example, Theorem \ref{tt} sheds light on the vanishing result obtained in Section \ref{mr0}. In this instance the manifold is indeed an anti-canonical hypersurface in a smooth ambient space. In addition, the sums of line bundles that appear in the definition of the bundle (\ref{Vdef}) do extend holomorphically to the ambient space. Although it is not guaranteed, it is also not unreasonable to think that the connection on this bundle might also extend holomorphically to the ambient space. Indeed, the ansatze that are used to describe fiber metrics in numerical work \cite{Donaldson,Headrick:2005ch,Douglas:2006hz,Douglas:2006rr,Braun:2007sn,Braun:2008jp,Headrick:2009jz,Anderson:2010ke,Anderson:2011ed,Ashmore:2019wzb,Cui:2019uhy} are somewhat suggestive of this. Given all of this, Theorem \ref{tt}, together with our discussion of Section \ref{mr0} makes it no surprise that $\CS_\omega(A)=0$.\\[2mm]
Despite the discussion of the proceeding paragraph, one might think that Theorem \ref{tt} has limited applicability in physical contexts. The techniques illustrated in the simple example in Section \ref{mr0} clearly generalize to large classes of cases and it might naively appear that the above theorem has a very limited scope in terms of the types of bundles and Calabi-Yau manifolds to which it applies. In fact, Theorem \ref{tt} is of relevance in a surprisingly wide range of examples.

The first seemingly strong restriction in Theorem \ref{tt} is the requirement that the Calabi-Yau manifold be described as an anti-canonical hypersurface in an ambient four-fold. This is in fact not much of a restriction at all in many discussions of string compactification. In the case of any complete intersection in a smooth ambient space, for example for any CICY \cite{Hubsch:1986ny,Candelas:1987kf,Green:1986ck,Candelas:1987du} or gCICY \cite{Anderson:2015iia} (see \cite{Berglund:2016yqo,Berglund:2016nvh} for related work), one can simply pick $Y$ to be described by $k-1$ of the defining equations where $k$ is the codimension of the three-fold. The final defining equation will then be an anti-canonical hypersurface in that ambient space. Further, for the theorem to apply it is only necessary that the Calabi-Yau manifold under consideration admits {\it some} description of this type. Calabi-Yau manifolds can be described in a plethora of different manners and even if a three-fold of interest is not described in a manner compatible with Theorem \ref{tt} that does not mean that such a description does not exist. Indeed, it can be hard in a given case to prove that a description as an anti-canonical hypersurface in an ambient $Y$ does not exist. 

It is true that many ambient spaces appearing in descriptions of known Calabi-Yau manifolds are singular. This is commonly the case for Calabi-Yau manifolds described as hypersurfaces in toric varieties \cite{Kreuzer:2000xy,Kreuzer:2002uu,Altman:2014bfa}, or quotients of CICYs \cite{Braun:2010vc,Candelas:2008wb,Candelas:2010ve,Candelas:2015amz,Candelas:2016fdy,Constantin:2016xlj,Candelas:2017ive}, for example. Even in such cases, however, the ambient spaces which appear in constructions in the literature are frequently resolvable and one can then simply apply Theorem \ref{tt} to the anti-canonical hypersurface in that resolution. If the initial Calabi-Yau manifold was smooth, then the ambient singularities must have missed the hypersurface and, hence, the three-fold is not changed during the resolution process.\\[2mm]
One might have similar reservations about the general applicability of the assumptions made about the bundles and connections as they appear in Theorem \ref{tt}. However, in this case too the structure required is not as restrictive as one might think and the theorem applies to many cases appearing in the physics literature. There are many constructions that are utilized in heterotic compactifications where the bundle does not extend nicely to the ambient space. Bundles constructed as two term monads over CICYs, for example, frequently have the feature that, while they restrict to a bundle over the Calabi-Yau three-fold they are merely a sheaf over the ambient space. As it happens, the bundle~(\ref{Vdef}) in our example extends to a bundle on the ambient space, but this is not necessarily the case in other models (see \cite{Anderson:2007nc,Anderson:2011ty} for some explicit cases). Nevertheless, even in these cases it is not clear that Thomas' theorem does not apply. Since the Chern-Simons invariant is unchanged under holomorphic deformations of its argument, we only really require some holomorphic deformation of the bundle under consideration to extend to the ambient space. In addition, as with manifolds, bundles over Calabi-Yau three-folds can typically be described in many ways. Even if some descriptions do not allow for an extension to an ambient space bundle there may well exist others which do.\\[2mm]
Using the tools presented in Section \ref{genproc}, we have computed the holomorphic Chern-Simons invariants for quite a number of different cases, and each time we have obtained zero. We believe that the above theorem may be one of the culprits behind this conspiracy. In the rest of this paper we will describe heterotic string compactifications in which non-trivial Chern-Simons invariants can be obtained, culminating in a concrete example of a non-flat gauge bundle giving rise to a fractional invariant.

\section{Fractional Chern-Simons invariants} \label{fracsec}

\subsection{General remarks}
In this section we will discuss two methods for constructing non-flat bundles in heterotic compactifications which give rise to fractional holomorphic Chern-Simons superpotential contributions. These discussions will focus on manifolds which are freely acting quotients of an initial simply connected Calabi-Yau three-fold (or on three-folds with a non-trivial fundamental group). The technical results that we will need as part of this discussion are presented in Appendix \ref{quotapp}.\\[2mm]
The first argument we wish to give makes use of large gauge transformations on a Calabi-Yau manifold $X$ in order to generate connections with fractional holomorphic Chern-Simons invariants on its quotient $\hat{X}$. Working over the geometry $X$, it is easy to obtain a vanishing Chern-Simons invariant. Indeed, an example of such a case is given in Section \ref{mr0}. From such a result one can easily obtain a non-vanishing, but non-fractional holomorphic Chern-Simons invariant, simply by performing a large gauge transformation on the argument of the functional, $A$. It is not clear what integers one can obtain for the associated ordinary Chern-Simons invariants in such a case, and for a given topology it is not the case that every possible integer will always be obtainable. Furthermore, explicitly writing down such large gauge transformations appears to be prohibitively difficult in many cases. Nevertheless, non-vanishing Chern-Simons invariants can clearly be obtained on $X$.\\[2mm]
Let us denote by $\Gamma$ the freely acting symmetry on $X$ by which we quotient to obtain $\hat{X}$. Further, a $\Gamma$-equivariant bundle $V$ on $X$ descends to a bundle on $\hat{X}$ which we denote by $\hat{V}$. In Appendix \ref{quotapp}, we introduce the notion of $\Gamma$-equivariant connections on $\Gamma$-equivariant bundles on $X$. Suppose we consider such $\Gamma$-equivariant connections $A$ and $A_0$ on $V$ which give rise to a holomorphic Chern-Simons invariant on $X$ with at least one of the numbers $a^i$ and $b_i$ in (\ref{WH2}) (that is, at least one of the ordinary Chern-Simons terms involved) not being divisible by $|\Gamma|$. In Appendix \ref{quotapp}, we show that the resulting holomorphic Chern-Simons invariant on $\hat{X}$ is obtained by dividing its counterpart on $X$ by the group order (see (\ref{upstairsdownstairs})). Therefore the holomorphic Chern-Simons invariant obtained on the quotient would be fractional.

One might think that such cases are common place, and indeed that may well be true. However, constructing a concrete example as an existence proof is difficult, due to the fact that an explicit expression for, or at least proof of existence of, the large gauge transformation involved is required. Without this one cannot concretely rule out the possibility that all large gauge transformations over $X$ that exist lead to integers divisible by $|\Gamma|$ for all possible symmetries by which the manifold could be quotiented, however unlikely this may seem.\\[2mm]
It is interesting to ask how this method for obtaining fractional holomorphic Chern-Simons invariants evades the statement that large gauge transformations on $\hat{X}$ should change those functionals by integer multiples of periods. Suppose we have an equivariant connection $A_0$ on the (equivariant) bundle $V\rightarrow X$, and another connection $A$ on $V$, related to $A_0$ by a large gauge transformation, so that $\CS_{A_0}(A)$ is an integer multiple of periods.  We prove in Appendix \ref{quotapp} that the large gauge transform of an equivariant gauge field is always equivariant so that $A$ is equivariant as well. Hence, both $A_0$ and $A$ descend to the quotient, inducing connections $\hat{A}_0$ and $\hat{A}$ of $\hat{V}$. However, it is not true that the large gauge transformation involved will always descend to $\hat{X}$. In other words, $\hat{A}_0$ and $\hat{A}$ need not be related by a large gauge transformation and, hence, the corresponding holomorphic Chern-Simons invariant $\CS_{\hat{A}_0}(\hat{A})$ on $\hat{X}$ does not have to be an integer multiple of periods.

Given the non-constructive nature of such an argument for the existence of fractional holomorphic Chern-Simons invariants, we will, in the following subsections, present a concrete example of a non-flat bundle exhibiting such a structure. We construct this example in a manner which is presumably less generic, but nevertheless more explicit, than the discussion of the proceeding paragraphs.

\subsection{Tensor product connections} 
To discuss the example in the next sub-section we will need a few basic facts and definitions concerning tensor product connections and their Chern-Simons invariants. Given two bundles $V_1$ and $V_2$ with connections $\nabla_1$ and $\nabla_2$, the tensor product $V=V_1\otimes V_2$, can be equipped with the tensor product connection $\nabla$ defined by
\begin{eqnarray} \label{tpdef}
\nabla (s_1 \otimes s_2) =( \nabla_1 s_1)\otimes s_2 + s_1 \otimes (\nabla_2 s_2)\; .
\end{eqnarray}
Here, $s_1$ and $s_2$ are sections of $V_1$ and $V_2$, respectively. If we set up local frames $s_{1i}$ and $s_{2k}$ for $V_1$ and $V_2$ then these define a local frame $s_{1i}\otimes s_{2k}$ for $V$. The corresponding gauge fields, introduced in the usual manner as
\begin{equation}
\nabla_1 s_{1\,i} = A_{1i}^{j} s_{1\,j}\; ,\qquad \nabla_2 s_{2\,k} = A_{2k}^{l} s_{2\,l}\;\qquad
\nabla(s_{1\,i}\otimes s_{2\,a}) = A_{i a}^{\,jb}(s_{1\,j}\otimes s_{2\,b})\; ,
\end{equation}
are then easily seen to be related by
\begin{equation}
A_{i k}^{\,jl} =A_{1i}^{j} \delta_k^l + \delta_i^j A_{2k}^{l}\; .
\end{equation}
For the curvature of the tensor product connection, it follows from the definition  (\ref{tpdef}) that
\begin{eqnarray}
F(s_1\otimes s_2)= (F_1( s_1))\otimes s_2 + s_1 \otimes (F_2( s_2))
\end{eqnarray}
Hence, if $F_1$ and $F_2$ satisfy the Hermitian Yang-Mills equations~(\ref{hym}) then so does $F$.\\[2mm]
Given this set-up, it is straightforward to compute the Chern-Simons form for the tensor product connection
\begin{eqnarray} \label{mrguy}
\omega_3(A) = \textnormal{rk}(V_2)\, \omega_3(A_1) + \textnormal{rk} (V_1)\, \omega_3(A_2) \;.
\end{eqnarray}
We introduce reference Chern-connections $A_{10}$ and $A_{20}$ for $A_1$ and $A_2$, respectively, along with their tensor product connection $A_0$ which serves as a reference connection for $A$. Then, (\ref{mrguy}) combined with the formula~(\ref{mhcs2}) for the holomorphic Chern-Simons invariant gives
\begin{eqnarray} \label{prodcs}
\CS_{A_0}(A) = \textnormal{rk} (V_2)\, \CS_{A_{10}}(A_1) + \textnormal{rk} (V_1)\, \CS_{A_{20}}(A_2)\; .
\end{eqnarray}

\subsection{A non-flat bundle with a fractional invariant} \label{finaleg}
We will now construct an example of a non-flat bundle with a fractional holomorphic Chern-Simons invariant as defined in Section \ref{sec:csi}, by using the notion of tensor product connections on a quotient of a CICY three-fold. It should be noted that finding calculable examples of this type is also rather difficult. The structure required, as we will see, is rather specific. In addition, most cases that both exhibit the necessary structure and are calculationally tractable have turned out not to lead to a fractional Chern-Simons invariant. Nevertheless, we find it valuable to provide this example as an existence proof for non-flat bundles in heterotic compactifications with fractional holomorphic Chern-Simons invariants.\\[2mm]
Consider the manifold, CICY 5301, in the standard list \cite{Candelas:1987kf,cicylist}, specified by the configuration matrix
\begin{eqnarray} \label{5301eg}
X\in \left[\begin{array}{c|cccc} \mathbb{P}^1& 0 &1&1&0 \\  \mathbb{P}^1& 0&1&1&0\\  \mathbb{P}^1& 1&0&0&1\\  \mathbb{P}^1&1&0&0&1\\ \mathbb{P}^3&1&1&1&1 \end{array} \right]\; .
\end{eqnarray}
We denote the homogeneous ambient space coordinates by $x_{a,i}$, where $a=1,\ldots ,5$ labels the projective factors and $i=0,1,\ldots $ its coordinates.

This manifold admits a freely acting $\mathbb{Z}_4$ symmetry whose generator acts as follows~\footnote{Note we have performed a linear coordinate transformation from the symmetry action as it is usually represented in the standard list \cite{Braun:2010vc,cicylist}. This form of the symmetry preserves a particularly simple SLAG as we will see shortly.}
\begin{equation}\label{sym}
\begin{array}{llrlllrlllllllr}
 x_{1,0}&\mapsto& x_{3,0}&&x_{1,1}&\mapsto& -x_{3,1}&&x_{2,0}&\mapsto& x_{4,0}&&x_{2,1}&\mapsto& x_{4,1}\\
 x_{3,0}&\mapsto& x_{1,0}&&x_{3,1}&\mapsto& x_{1,1}&&x_{4,0}&\mapsto&x_{2,0}&&x_{4,1}&\mapsto& -x_{2,1}\\
 x_{5,0}&\mapsto&x_{5,3}&&x_{5,1}&\mapsto& -x_{5,2}&&x_{5,2}&\mapsto& x_{5,1}&&x_{5,3}&\mapsto& x_{5,0}
\end{array}\; .
\end{equation}
The symmetry also acts non-trivially on the normal bundle as represented by the following action
\begin{eqnarray} \label{sym2}
\left(p_1,p_2,p_3,p_4 \right) \mapsto \left(p_2,p_1,p_4,-p_3 \right),
\end{eqnarray}
on the defining polynomials. The quotient $\hat{X}$ of $X$ by the symmetry (\ref{sym}), (\ref{sym2}) leads to a transverse variety, and the action is fixed point free, and, hence, $\hat{X}$ is a smooth Calabi-Yau three-fold with fundamental group $\pi_1(\hat{X})= \mathbb{Z}_4$. It is on $\hat{X}$ that we will construct our example.\\[2mm]
To construct our bundle we begin by noting that the following sum of line bundles \begin{eqnarray}
U = {\cal O}_X(-2,-1,0,0,1) \oplus {\cal O}_X(0,1,-2,1,0) \oplus {\cal O}_X(2,0,2,-1,-1)
\end{eqnarray}
has a second Chern character which is exactly half of that of the tangent bundle $TX$. In addition, $U$ has a vanishing slope on an appropriate sub-locus of the K\"ahler cone and admits an equivariant structure with respect to the symmetry (\ref{sym}), (\ref{sym2}), so that it descends to a bundle $\hat{U}$ on $\hat{X}$. Now consider the bundle $V=U \oplus U$ on $X$. This is an equivariant bundle with a second Chern character which matches that of $TX$ and thus naively gives a good heterotic vacuum on $X$. Its equivariant nature means that it descends to a bundle $\hat{V} = \hat{U} \oplus \hat{U}$ on $\hat{X}$. We also introduce the well-know holomorphic deformation of $TX \oplus {\cal O}_X^{\oplus 5}$, given by
\begin{eqnarray}  \nonumber
0 \to V_0 \to {\cal O}_X(1,0,0,0,0)^{\oplus 2} \oplus \ldots \oplus {\cal O}_X(0,0,0,0,1)^{\oplus 4} \to {\cal O}_X(0,0,1,1,1)^{\oplus 2} \oplus {\cal O}_X(1,1,0,0,1)^{\oplus 2} \to 0
\end{eqnarray}
This bundle also has an equivariant structure under the above $\mathbb{Z}_4$ symmetry and descends to a bundle $\hat{V}_0$ on $\hat{X}$. For all these bundles, we introduce Chern connections which satisfy the Hermitian Yang-Mills equations, as indicated in Table~\ref{tab:bundles}. Note that the connection $\hat{A}$ on $\hat{V}$ is taken as a direct sum connection constructed from two copies of the connection $A_{\hat{U}}$ on $\hat{U}$. 
\begin{table}[!h]
\begin{center}
\begin{tabular}{|c||c|c|c||c|c|c|c|c|}\hline
space&\multicolumn{3}{c||}{$X$}&\multicolumn{5}{c|}{$\hat{X}=X/\mathbb{Z}_4$}\\\hline
bundle&$U$&$V=U^{\oplus 2}$&$V_0$&$\hat{U}$&$\hat{V}=\hat{U}^{\oplus 2}$&$\hat{V}_0$&$\hat{W}$&$\hat{V}'=\hat{U}\oplus\hat{W}$\\\hline
connection&$A_U$&$A$&$A_0$&$A_{\hat{U}}$&$\hat{A}$&$\hat{A}_0$&$A_{\hat{W}}$&$\hat{A}'$\\\hline
\end{tabular}
\caption{Bundles and associated connections of our construction}\label{tab:bundles}
\end{center}
\end{table}
\vskip -5mm\noindent
We are interested in the holomorphic Chern-Simons invariant $\CS_{\hat{A}_0}(\hat{A})$. If this invariant is fractional then we can stop our search here. If it is not, however, there is a simple modification that allows us to generate a new bundle on the quotient manifold $\hat{X}$ which does have a fractional invariant. More specifically, we can can carry out the following modification
\begin{equation}
  \hat{V}= \hat{U} \oplus \hat{U} = \hat{U} \otimes ({\cal O}_{\hat{X}} \oplus {\cal O}_{\hat{X}})\quad\longrightarrow\quad \hat{V}' =\hat{U}\otimes \hat{W}
\end{equation} 
of the bundle $\hat{V}$, where $\hat{W}$ is a rank two flat bundle (a Wilson line) on $\hat{X}$ which we have used to replace the trivial bundle ${\cal O}_{\hat{X}} \oplus {\cal O}_{\hat{X}}$.
In doing so we do not change the second Chern-character of the bundle. In addition, if we choose the connection $\hat{A}'$ on $\hat{V}'$ to be the tensor product connection of the Hermitian-Yang-Mills connection $A_{\hat{U}}$ on $\hat{U}$ and the flat connection $A_{\hat{W}}$ on $\hat{W}$ then the result will still obey the Hermitian Yang-Mills equations (\ref{hym}). Thus the resulting bundle $\hat{V}'$ still gives rise to a good heterotic vacuum before considering Chern-Simons contributions to the superpotential.\\[2mm]
We would now like to compare the holomorphic Chern-Simons invariants $\CS_{\hat{A}_0}(\hat{A})$ and $\CS_{\hat{A}_0}(\hat{A'})$ for $\hat{A}$ and $\hat{A}'$, relative to the same reference connection $\hat{A}_0$. From (\ref{mhcs2}) we know that
\begin{equation}
\CS_{\hat{A}_0}(\hat{A}) = \int \left( \omega_3(\hat{A}) - \omega_3(\hat{A}_0) \right) \wedge \hat{\Omega}= \int \left(2 \omega_3(A_{\hat{U}}) - \omega_3(\hat{A}_0) \right) \wedge \hat{\Omega}
\end{equation} 
Here $\hat{\Omega}$ is the holomorphic $(3,0)$-form on the quotient $\hat{X}$. Of course, appropriate morphisms must be used to ensure that the two connections in the above integral are written with respect to the same trivialization on the underlying real bundle.

On the other hand, for the connection $\hat{A}'$, we make use of the formula (\ref{mrguy}) which gives
\begin{equation}
\omega_3(\hat{A}') = 2\, \omega_3(A_{\hat{U}}) + 3\, \omega_3(A_{\hat{W}})\; .
\end{equation}
This, in turn, leads to the following relation
\begin{equation} \label{CSAVp}
\CS_{\hat{A}_0}(\hat{A}')= \int \left(2\, \omega_3(A_{\hat{U}}) +3\, \omega_3(A_{\hat{W}}) - \omega_3(\hat{A}_0) \right) \wedge \hat{\Omega}=
\CS_{\hat{A}_0}(\hat{A}) +3\, \CS_0(A_{\hat{W}})\; ,
\end{equation}
where $\CS_0(A_{\hat{W}})$ denotes the holomorphic Chern-Simons invariant with a trivial reference connection, which is always available on flat bundles.
Thus, under our assumption that $\CS_{\hat{A}_0}(A_{\hat{V}})$ is non-fractional, a flat bundle $\hat{W}$ with a non-integer value for $3\, \CS_0(A_{\hat{W}})$, leads to a fractional Chern-Simons invariant for $\hat{A}'$. The study of Chern-Simons invariants of flat bundles of this type is more advanced in the heterotic literature than for their non-flat cousins \cite{Conrad:2000tk,Gukov:2003cy,Apruzzi:2014dza}. The above construction allows us to use these methods for flat bundles in order to construct non-flat bundles with fractional Chern-Simons invariants.\\[2mm]
In order to show that $3\CS_0(A_{\hat{W}})$ can be non-integer, it is enough to find a special Lagrangian three-cycle (SLAG) ${\cal C}$ such that the ordinary Chern-Simons invariant $3\, \textnormal{OCS}_0(A_{\hat{W}},{\cal C})$ is non-integer. In the following we follow the analysis and notation of Ref.~\cite{Apruzzi:2014dza}. It turns out, the Calabi-Yau three-fold $X$ described by the configuration matrix~(\ref{5301eg}) admits an A-type SLAG which can be written as the configuration
\begin{eqnarray} \label{mrslag}
{\cal C}\in\left[\begin{array}{c|cccc} \mathbb{RP}^1& 0 &1&1&0 \\  \mathbb{RP}^1& 0&1&1&0\\  \mathbb{RP}^1& 1&0&0&1\\  \mathbb{RP}^1&1&0&0&1\\ \mathbb{RP}^3&1&1&1&1 \end{array} \right]\;.
\end{eqnarray}
One can always solve the four equations given here to obtain a single point in $\mathbb{RP}^1 \times\mathbb{RP}^1 \times\mathbb{RP}^1 \times\mathbb{RP}^1$. Perhaps the easiest way to see this is to consider (\ref{5301eg}) as a point fibration over $\mathbb{P}^3$. That point fibration does degenerate, of course, but the degeneracy locus misses the SLAG.  Hence, the above configuration is simply a description of the Lens space $\mathbb{RP}^3 = S^3 /\mathbb{Z}_2$.

It is clear that the $\mathbb{Z}_4$ symmetry (\ref{sym}) and (\ref{sym2}) leaves the SLAG ${\cal C}$ invariant and, in the quotient $\hat{X}$, it turns into the lense space $\hat{\cal C}=S^3/\mathbb{Z}_8$. Flat bundles are defined uniquely by a map from the fundamental group of the base space to the structure group. As such, we can easily see how a flat bundle defined on the whole Calabi-Yau three-fold restricts to a SLAG simply by looking at how non-trivial one-cycles embedded in the SLAG descend from the ambient manifold. 

For a Lens space with fundamental group is $\mathbb{Z}_p$ we can define an $SU(N)$ flat bundle by specifying the images of the map  $\mathbb{Z}_p\rightarrow SU(N)$ and we denote these by $\diag(e^{2\pi i k_1/p},\ldots, e^{2 \pi i k_N/p})$ , where $k_i\in\mathbb{Z}$. Then the general formula for the Chern-Simons invariant of such a flat bundle, using the trivial connection as the reference, is as follows \cite{Witten:1985mj,kirkklassen,Rozansky:1993zx,Nishi}.
\begin{eqnarray} \label{lenscs}
\textnormal{OCS}_0(A_{\hat{W}},S^3/\mathbb{Z}_p) = -\sum_i \frac{k_i^2}{2p} \mod \mathbb{Z}
\end{eqnarray}
For the case at hand, we have $p=8$ and we choose the images of the defining map $\mathbb{Z}_4\rightarrow SU(2)$ as ${\rm diag}(e^{2 \pi i  2/8} , e^{ 2 \pi i  6/8})$. This restricts to the Lens space in an obvious manner. The ordinary Chern-Simons invariant evaluated on this Lens space with the restricted flat bundle is then
\begin{equation}
 \textnormal{OCS}_0(A_{\hat{W}},\hat{\cal C})=\frac{5}{2}\;{\rm mod}\;\mathbb{Z}\; .
\end{equation}
As a result $3\, \textnormal{OCS}_0(A_{\hat{W}},{\cal C})$ is non-integer and, hence, from (\ref{CSAVp}), either $\CS_{\hat{A}_0}(\hat{A})$ or  $\CS_{\hat{A}_0}(\hat{A}')$ is fractional. In conclusions, we have obtained a contribution to a heterotic superpotential from a fractional holomorphic Chern-Simons invariant associated to a non-flat bundle.\\[2mm]
There is an important caveat in the above example that should be mentioned. Although it is true that second Chern-characters of $\hat{V}$ and $\hat{V}'$ match at the level of the image of the Chern-Weil homomorphism, it is not clear they match in torsion. This is also required for a viable heterotic vacuum, and indeed for the holomorphic Chern-Simons invariant (\ref{mhcs}) to be well defined. The Brauer group of the manifold $\hat{X}$ is, to our knowledge, unknown and its computation is beyond the scope of this paper. This is unfortunately, a common situation in heterotic compactifications. Nevertheless we believe the present example exemplifies well the idea of the construction. 


\section{Conclusions and outlook} \label{conc}

In this paper we have computed the Chern-Simons contribution to the heterotic superpotential arising from the interplay between the gauge and the tangent bundles. To do this we have split the superpotential which originates from the NS fields strength $H$ up into two pieces, one from harmonic flux which is integer quantized and the other from the Chern-Simons invariant. The second contribution is potentially fractional and has been the main focus of the present work.   Alternative, we might say that the main purpose of this paper has been to determine the quantization condition for $H$. From this point of view, bundles with fractional Chern-Simons invariants do not allow for a vanishing $H$ and, hence, lead to a non-zero flux superpotential.\\[2mm]
Chern-Simons invariants in the context of heterotic string compactifications have been considered previously, but only in the context of flat (Wilson line) bundles. However, heterotic compactifications on Calabi-Yau manifolds require non-flat gauge bundles and it is, therefore, essential to analyze Chern-Simons invariants for such cases. In the present paper, we have presented the first analysis of this kind.\\[2mm]
We have developed a number of new methods to carry out our computations. Explicit real bundle isomorphisms between line bundle sums on $\mathbb{P}^1$ have been derived and we have shown how these isomorphisms can be used to construct real bundle isomorphisms between line bundle sums on Calabi-Yau manifolds which are defined in ambient spaces with $\mathbb{P}^1$ factors. These isomorphisms, together with holomorphic deformations, can be combined to isomorphisms between the tangent bundle of the Calabi-Yau manifold and the heterotic gauge bundle. This in turn allows for an explicit calculation of the gauge bundle's Chern-Simons invariant, with the tangent bundle as the reference connection.

Further, we have developed methods for calculating Chern-Simons invariants on Calabi-Yau quotient manifolds, that is, on manifolds with a non-trivial first fundamental group, which apply to non-flat bundles. Since realistic heterotic Calabi-Yau compactifications rely on such a quotient constructions for both the manifold and the bundle, these methods are essential for analyzing the superpotential for phenomenologically relevant models.\\[2mm]
Using the methods based on real bundle morphisms, we have calculated the holomorphic Chern-Simons invariants for many examples and have always found a vanishing result. Presumably many of these results can be attributed to the vanishing theorem~\ref{tt}. A non-zero Chern-Simons invariant causes a large superpotential contribution which, on it own, de-stabilizes the model, so the frequent vanishing we have found can be considered good news.
However, we have also presented an example of a non-zero and indeed fractional Chern-Simons invariant for a non-flat bundle on a quotient Calabi-Yau.\\[2mm]
Knowledge of Chern-Simons superpotentials in heterotic theories is a crucial piece of information, particularly in view of vacuum stability and moduli stabilization. In this paper, we have presented some progress in calculating such superpotentials but much remains to be done for a systematic understanding of heterotic vacua. To generalize our methods to larger classes of models more general real bundle isomorphisms need to be constructed. So far, our approach is based on rank two line bundle sums on $\mathbb{P}^1$. Explicit knowledge of bundle isomorphisms for higher rank line bundle sums on $\mathbb{P}^1$ and for higher-dimensional projective spaces would significantly expand the scope for applications. Other methods to construct real bundle isomorphisms, for example through deformations to exceptional structure groups such as $G_2$, might also be of interest (or perhaps methods taking a complementary geometric approach \cite{clemens_thomas}). One long term goal of this work is to derive a general quantization rule for $H$. Such a rule would be a potentially powerful model-building tool, and would allow us to distinguish marginally stable from unstable heterotic models.

Another obvious extension of the present work would be to include cases where five-branes are present in the vacuum. This would require new mathematics in that a suitable  generalization of Chern-Simons invariants would have to be formulated. This would certainly be interesting to pursue, and is perhaps a case where physics could guide the discovery of new mathematical structures.

It would also be interesting to study the effects which lead to these Chern-Simons superpotentials in dual theories, such as, for example, F-theory models with heterotic duals~\cite{Anderson:2010mh,Anderson:2011ty,Anderson:2013qca,Anderson:2011cza}. The authors are planning to explore some of these directions in future publications.


\section*{Acknowledgments}

The authors would like to thank Michael Albanese, Claude LeBrun, and Laura Schaposnik for very useful conversations. The work of L.A, J.G. and J.W. are supported in part by NSF grant PHY-1720321. The authors would like to gratefully acknowledge the hospitality of the Simons Center for Geometry and Physics (and the semester long program, The Geometry and Physics of Hitchin Systems) where work on this research was begun.


\appendix

\section{Real bundle morphisms} \label{realbunmorpheg}
In this appendix, we review some standard mathematics concerning bundles and their morphisms and we construct explicitly the real bundle isomorphisms between rank two line bundle sums on $\mathbb{P}^1$ which are used in the main part of the paper.\\[2mm]
We start by recalling some general facts about bundle morphisms. Suppose we have two bundles
\begin{equation}
 V\stackrel{\pi}{\longrightarrow} X\;,\qquad \tilde{V}\stackrel{\tilde{\pi}}{\longrightarrow} X\; ,
\end{equation} 
with typical fiber $F$ over a manifold $X$ with cover $U_\alpha$ and charts $\varphi_\alpha:U_\alpha\rightarrow W_\alpha\subset\mathbb{C}^n$. A bundle morphism is a map $f:V\rightarrow \tilde{V}$, for which the diagram
\begin{equation}
\begin{array}{lllll}
 &V&\stackrel{f}{\longrightarrow}& \tilde{V}&\\
 \pi&\downarrow&&\downarrow&\tilde{\pi}\\
&X&\stackrel{\rm id}{\longrightarrow}&X&
\end{array}
\end{equation}
commutes. We are looking for a practical way to construct such bundle morphisms and to this end we introduce local trivializations and their associated transition functions
\begin{equation}
\begin{array}{clllcll}
 \phi_\alpha&:&\pi^{-1}(U_\alpha)\rightarrow W_\alpha\times F&\qquad&  \tilde{\phi}_\alpha&:&\tilde{\pi}^{-1}(U_\alpha)\rightarrow W_\alpha\times F\\
 \phi_{\alpha\beta}:=\phi_\alpha\circ\phi_\beta^{-1}&:&W_\beta\times F\rightarrow W_\alpha\times F&\qquad&
 \tilde{\phi}_{\alpha\beta}:=\tilde{\phi}_\alpha\circ\tilde{\phi}_\beta^{-1}&:&W_\beta\times F\rightarrow W_\alpha\times F
\end{array} \; .
\end{equation} 
Given this set-up, we can define local versions of the bundle morphism $f$ by
\begin{equation}
 f_\alpha:=\tilde{\phi}_\alpha\circ f\circ\phi_\alpha^{-1}:W_\alpha\times F\rightarrow W_\alpha\times F\; ,
\end{equation} 
and a simple calculation shows that these local morphisms have to satisfy the intertwining rules
\begin{equation}\label{fcond}
f_\alpha\circ\phi_{\alpha\beta}= \tilde{\phi}_{\alpha\beta}\circ f_\beta\; , 
\end{equation} 
on the overlaps $(W_\alpha\cap W_\beta)\times F$. Conversely, any collection of local morphisms $f_\alpha$ which satisfies the  conditions~(\ref{fcond}) defines a bundle morphism $f$. To be more explicit, we introduce coordinates $(z,v)\in W_\beta\times F$ and write the transition functions and local morphisms as
\begin{equation}
 \phi_{\alpha\beta}(z,v)=\left(z,T_{\alpha\beta}(z)v\right)\;,\qquad
 \tilde{\phi}_{\alpha\beta}(z,v)=(z,\tilde{T}_{\alpha\beta}(z)v)\;,\qquad
 f_\beta(z,v)=\left(z,P_\alpha(z,\bar{z})v\right)\;,
\end{equation} 
where $T_{\alpha\beta}$, $\tilde{T}_{\alpha\beta}$ and $P_\alpha$ are $z$-dependent matrices which act on the fiber. Using this notation, the intertwining conditions~(\ref{fcond}) translate into the matrix equations
\begin{equation}
P_{\alpha} T_{\alpha\beta}= \tilde{T}_{\alpha\beta}P_\beta\; . \label{fcond1}
\end{equation} 
These conditions point to a practical way of finding bundle morphisms. Suppose we are given the transition functions $T_{\alpha\beta}$ and $\tilde{T}_{\alpha\beta}$ for the two bundles $V$ and $\tilde{V}$. Then, the task is to find matrices  $P_\alpha$ which contain smooth functions on $W_\alpha$, are invertible for all $z\in W_\alpha$ and satisfy the matrix relations~(\ref{fcond1}). These matrices then define a bundle isomorphism $f\sim(P_\alpha):V\rightarrow\tilde{V}$ which establishes the equivalence of the two bundles.\\[2mm]
We will now apply this method to find isomorphisms between line bundle sums on $X=\mathbb{P}^1$. The two standard patches on $\mathbb{P}^1$ are denoted by $U_0\cong\mathbb{C}$ and $U_1\cong\mathbb{C}$, with affine coordinates $z\in U_0$ and $w=1/z\in U_1$. Line bundles on $\mathbb{P}^1$ are denoted by ${\cal O}_{\mathbb{P}^1}(k)$, as usual.

It is known that two line bundle sums on $\mathbb{P}^1$ with the same rank are (real) isomorphic if their first Chern classes match. Our task is to construct this isomorphism explicitly for the case of rank two line bundle sums
\begin{equation}
 V(k,l):={\cal O}_{\mathbb{P}^1}(k)\oplus{\cal O}_{\mathbb{P}^1}(l)\; ,\qquad T^{(k,l)}_{10}={\rm diag}(z^{-k},z^{-l})\; ,
\end{equation} 
with  transition functions $T^{(k,l)}_{10}$. We start by considering the two bundles $V=V(-p,p)$ and $\tilde{V}=V(0,0)$ where $p>0$. Evidently, they are both rank two bundles with vanishing first Chern class so they must be a real isomorphism $f^{(p)}\sim (P^{(p)}_0,P^{(p)}_1):V(-p,p)\rightarrow V(0,0)$. To find this isomorphism explicitly, we write down the transition functions
\begin{equation}
T_{10}=T_{10}^{(-p,p)}(z)={\rm diag}(z^p,z^{-p})\;,\qquad \tilde{T}_{10}=T_{10}^{(0,0)}(z)={\rm diag}(1,1)\; ,
\end{equation}
and we try to find non-singular matrices $P^{(p)}_\alpha$ which satisfy the intertwining conditions (\ref{fcond1}). For the present case, we have only two patches so there is only one such condition which reads
\begin{equation}
 P_1^{(p)}T_{10}^{(-p,p)}=T_{10}^{(0,0)}P_0^{(p)}\; . \label{fcondP1}
\end{equation}
Here $P_0^{(p)}$ contains smooth functions in $z\in U_0\cong \mathbb{C}$ and is invertible everywhere in its domain and $P_1^{(p)}$ contains smooth functions in $w\in U_1\cong\mathbb{C}$ and is also invertible everywhere in its domain. Starting with a guess for $P_1^{(p)}$, (\ref{fcondP1}) then determines $P_0^{(p)}$ and this leads to
\begin{equation}
  P_1^{(p)}(w,\bar{w})=\left(\begin{array}{cc}w^p&\frac{1}{1+|w|^{2p}}\\-1&\frac{\bar{w}^p}{1+|w|^{2p}}\end{array}\right)\quad\Longrightarrow\quad
  P_0^{(p)}(z,\bar{z})=\left(\begin{array}{cc}1&\frac{\bar{z}^p}{1+|z|^{2p}}\\-z^p&\frac{1}{1+|z|^{2p}}\end{array}\right)\; . \label{P01def}
\end{equation}  
Evidently, both matrices are smooth in their respective domain, they are invertible since ${\rm det}(P_0^{(p)})={\rm det}(P_1^{(p)})=1$ for all $z,w\in\mathbb{C}$ and they satisfy (\ref{fcondP1}) by construction. So, in conclusion, this defines real bundle isomorphisms
\begin{equation}
 f^{(p)}\sim\left(P_\alpha^{(p)}\right):V(-p,p)\stackrel{\simeq}{\longrightarrow} V(0,0)\; .
\end{equation} 
As mentioned above, it is a well-known fact that these bundles are real isomorphic~\cite{okonek} (see \cite{Douglas:2004yv} for a discussion in the physics literature). However, their isomorphy is normally established in a somewhat different manner and we are not aware of the explicit real isomorphism being written down in this form in the literature. It is this kind of construction that we will need in the rest of the paper, hence the above discussion.\\[2mm]
The above construction can easily be generalized by twisting up with another line bundle. The transitions function for the bundle $V(a-p,a+p)=V(-p,p)\otimes {\cal O}_{\mathbb{P}^1}(a)$ satisfies
\begin{equation} \label{tensortrans}
 T_{10}^{(a-p,a+p)}=z^{-a}T_{10}^{(-p,p)}\; .
\end{equation} 
Hence, multiplying (\ref{fcondP1}) with $z^{-a}$ it follows easily that
\begin{equation}
 P_1^{(p)}T_{10}^{(a-p,a+p)}=T_{10}^{(a,a)}P_0^{(p)}\; , \label{fcondP1t}
\end{equation}
for the same matrices $P_0^{(p)}$ and $P_1^{(p)}$ as given in (\ref{P01def}). Hence, we have explicitly constructed the real bundle isomorphisms
\begin{equation}
  f^{(p)}\sim\left(P_\alpha^{(p)}\right):V(a-p,a+p)\stackrel{\simeq}{\longrightarrow} V(a,a)\; .
\end{equation} 
between two rank two line bundle sums on $\mathbb{P}^1$ with the same {\it even} first Chern class.\\[2mm]
What about the case of two rank two line bundle sums with the same {\it odd} first Chern class? Define the matrix $D={\rm diag}(1,z)$ and multiply (\ref{fcondP1t}) with this matrix from the right. This leads to
\begin{equation}
 Q_1^{(p)}T_{10}^{(a-p,a+p+1)}=T_{10}^{(a,a+1)}Q_0^{(p)}\; , \label{fcondQ1t}
\end{equation}
where
\begin{equation}
 Q_1^{(p)}=P_1^{(p)}\;,\quad Q_0^{(p)}=D^{-1}P_0^{(p)}D=\left(\begin{array}{cc}1&\frac{z\bar{z}^p}{1+|z|^{2p}}\\-z^{p-1}&\frac{1}{1+|z|^{2p}}\end{array}\right)\; .
\end{equation} 
Note that $Q_0^{(p)}$ and $Q_1^{(p)}$ are still smooth in their respective coordinates and ${\rm det}(Q_0^{(p)})={\rm det}(Q_1^{(p)})=1$ for all $z,w\in\mathbb{C}$. This means we have the real bundle isomorphisms
\begin{equation}
 g^{(p)}\sim\left(Q_\alpha^{(p)}\right):V(a-p,a+p+1)\stackrel{\simeq}{\longrightarrow} V(a,a+1)\;.
\end{equation} 
between two  rank two line bundle sums on $\mathbb{P}^1$ with the same odd first Chern.\\[2mm]
So far, we have constructed isomorphism to somewhat special bundles of the form $V(a,a)$ or $V(a,a+1)$. This limitation is easily removed by introducing the matrices
\begin{equation} \label{mrP}
 P^{(q,p)}_\alpha(z,\bar{z}):={P^{(q)}_\alpha(z,\bar{z})}^{-1}P^{(p)}_\alpha(z,\bar{z})\;,\quad
 Q^{(q,p)}_\alpha(z,\bar{z}):={Q^{(q)}_\alpha(z,\bar{z})}^{-1}Q^{(p)}_\alpha(z,\bar{z})\; .
\end{equation} 
Note that these matrices are still well-defined on their respective patches - since we are dealing with ${\rm SL}(2,\mathbb{C})$ matrices the inverse does not introduce any singularities. By transitivity, these matrices satisfy
\begin{equation}
P_\alpha^{(q,p)} T_{\alpha\beta}^{(a-p,a+p)}=T_{\alpha\beta}^{(a-q,a+q)}P_\beta^{(q,p)}\;,\qquad
Q_\alpha^{(q,p)} T_{\alpha\beta}^{(a-p,a+p+1)}=T_{\alpha\beta}^{(a-q,a+q+1)}Q_\beta^{(q,p)}\; .
\end{equation}
and, hence, they define bundle isomorphisms
\begin{eqnarray} \label{fmorph}
 f^{(q,p)}&\sim&\left(P^{(q,p)}_\alpha\right):V(a-p,a+p)\stackrel{\simeq}{\longrightarrow} V(a-q,a+q)\\
 g^{(q,p)}&\sim&\left(Q^{(q,p)}_\alpha\right): V(a-p,a+p+1)\stackrel{\simeq}{\longrightarrow} V(a-q,a+q+1)
\end{eqnarray} 
between two arbitrary rank two line bundle sums on $\mathbb{P}^1$ with even and odd first Chern class, respectively. \\[2mm]
In conclusion, we have shown explicitly, by writing down the relevant real bundle isomorphisms, the well known fact that the first Chern class really does classify rank two line bundle sums on $\mathbb{P}^1$ as topological bundles. Crucially we have explicit forms for the relevant isomorphisms which will be important to us in the main part of this paper.

\section{Quotients and equivariant structures} \label{quotapp}

\vspace{0.2cm}

We require a small amount of mathematical formalism in order to describe the relationship between holomorphic Chern-Simons invariants on Calabi-Yau three-folds $X$ and their quotients by freely acting symmetries $\hat{X}$.\\[2mm]
{\bf Calabi-Yau quotients and equivariant bundles:} Let us introduce a finite group $\Gamma=\{g_0=e,g_1,\ldots ,g_n\}$, where $n=|\Gamma|-1$, which acts freely on the Calabi-Yau three-fold $X$. The quotient by this symmetry is denoted $\hat{X}=X/\Gamma$ and $p:X\rightarrow \hat{X}$ is the natural projection to the quotient.  We need to define the notion of a $\Gamma$-equivariant vector bundle $V\rightarrow X$. A bundle is equivariant if there exists bundle morphisms $\Phi_g$, for all $g\in\Gamma$, such that the diagrams
\begin{equation}
 \begin{array}{rcccl}
 &V&\stackrel{\Phi_g}{\longrightarrow}&V&\\
\pi&\downarrow&&\downarrow&\pi\\
 &X&\stackrel{g}{\longrightarrow}&X&
\end{array}
\end{equation} 
commute and the bundle morphisms satisfy the group law $\Phi_{gh}=\Phi_g\circ\Phi_h$ for all $g,h\in \Gamma$. Such a $\Gamma$-equivariant bundle descends to a bundle $\hat{V}\rightarrow\hat{X}$ on the quotient such that $V=p^*\hat{V}$. More constructively, the downstairs bundle can be defined as $\hat{V}=V/\sim$, with the equivalence relation defined by $v'\sim v\Leftrightarrow v'=\Phi_g(v)$ for a $g\in\Gamma$. The downstairs projection can be defined as $\hat{\pi}([v]):=[\pi(v)]$.

We also recall that the bundle morphisms $\Phi_g$ can be used to defined maps $\Psi_g:\Gamma(X,V)\rightarrow\Gamma(X,V)$ between sections of $V$ by
\begin{equation} \label{psig}
 \Psi_g(s):=\Phi_g\circ s\circ g^{-1}\; .
\end{equation} 
\vspace{0.1cm}

\noindent{\bf Interplay of equivariant structure and real bundle morphisms:} The possible choices of equivariant structure on holomorphic bundles $V$ on $X$ are in 1-1 correspondence with holomorphic bundles $\hat{V}$ on $\hat{X}$. Because of this there is a compatibility requirement between the possible real isomorphisms $f$ between two bundles $V$ and $\tilde{V}$ and choices of equivariant structure on each, if a real bundle isomorphism is to descend to the quotient. If two equivariant structures are to give rise to real isomorphic bundles on $\hat{X}$ then there should exist {\it some} real bundle isomorphism $f$ for the choices of equivariant structure $\Phi_g$ and $\tilde{\Phi}_g$ such that the following diagram commutes for all $g\in \Gamma$. 
\begin{equation} \label{comfp}
 \begin{array}{rcccl}
 &V&\stackrel{f}{\longrightarrow}&\tilde{V}&\\
 \Phi_g&\downarrow&&\downarrow&\tilde{\Phi}_g\\
 &V&\stackrel{f}{\longrightarrow}&\tilde{V}&
\end{array}
\end{equation} 
Note that any given real isomorphism $f:V\to\tilde{V}$ may not satisfy this condition. The requirement, if the two bundles on $\hat{X}$ are to be real isomorphic, is simply that there exists some bundle morphism that does.\\[2mm]
{\bf Pullback of a connection:}
In the context of this paper we need to go beyond the bundles themselves and consider also connections on them.  In order to consider how maps such as $\Phi_g$ or $f$ induce a mapping on connections we will need to recall how pullbacks of such objects can be defined. Consider an invertable differentiable map $f:X\rightarrow \hat{X}$ between two manifolds $X$, $\hat{X}$ and a vector bundle $\hat{V}\rightarrow \hat{X}$ with connection $\hat{\nabla}$. Then a connection $f^*\hat{\nabla}$ on the pullback bundle $V=f^*\hat{V}\rightarrow X$ can be defined locally as follows \cite{huybrechts}
\begin{equation}
 \nabla_i^{(0)}:=f^*\hat{\nabla}|_{U_i}:=d+f^*\hat{A}_i\; . \label{localconn}
\end{equation}
Here, $\hat{A}_i$ is the gauge field on $\hat{U}_i$ relative to the frames $\hat{s}_{i,a}$ on the cover $\hat{U}_i\subset\hat{X}$ and $U_i=f^{-1}(\hat{U}_i)$. There are natural frames on $U_i\subset X$ defined by $s_{i,a}=\hat{s}_{i,a}\circ f$, and we can glue the local connections $\nabla_i^{(0)}$ together to a global connection $\nabla^{(0)}$ relative to these frames. To see how this works we first note that gauge transformations
\begin{equation}
 A_g:=g^{-1}Ag+g^{-1}dg\;,
\end{equation} 
and pull-backs $f^*A$ commute, that is,
\begin{equation}
 f^*A_g=(f^*A)_{f^*g}\; .
\end{equation} 
This means, if the gauge fields  $\hat{A}_j$ and $\hat{A}_i$ on $\hat{U}_j$ and $\hat{U}_i$ are related by the gauge transformation $\hat{A}_j=\hat{A}_{i,\hat{g}_{(ij)}}$, then
\begin{equation}
 f^*\hat{A}_j=(f^*\hat{A}_i)_{g_{(ij)}}\; , \label{glue1}
\end{equation} 
so the pulled-back gauge fields $f^*\hat{A}_j$ and $f^*\hat{A}_i$ are glued together by the pull-backs $g_{(ij)}=f^*\hat{g}_{(ij)}$ of the original gauge transformations.

The above is natural but is not the most general way we can define the pullback. Suppose that, instead of (\ref{localconn}), we define local connections $\nabla_i$ on $U_i$ by
\begin{equation} 
 \nabla_i=d+A_i\;,\qquad f^*\hat{A}_i=A_{i,P_i}=P_i^{-1}A_iP_i+P_i^{-1}dP_i\; , \label{APdef}
\end{equation}
where $P_i$ are gauge transformations which can be chosen and the second equation defines what we mean by $A_i$. These local connections glue together to a global connection $\nabla$ by virtue of
\begin{equation}
 A_j=A_{i,P_ig_{(ij)}P_j^{-1}}
\end{equation}
(which follows immediately by combining (\ref{glue1}) and (\ref{APdef})), so the glueing gauge transformations in this case are given by $P_ig_{(ij)} P_j^{-1}$. Of course the so-defined gauge field $A_i$ depends on the choice of the local gauge transformations $P_i$. These different gauge fields describe the same pullback connection for different choices of local trivializations.\\[2mm]
{\bf Equivariant connections:} We will say that a connection $\nabla$ on $V\rightarrow X$ is $\Gamma$-equivariant\footnote{See \cite{bradlow_schaposnik} for related definitions in a different geometric context.} (with respect to the equivariant structure on $V$ defined by bundle morphisms $\Phi_g$) iff
\begin{equation}
 \nabla(\Psi_g(s))=\Psi_g'(\nabla(s))\; , \label{equivcond}
\end{equation}
for all $g\in\Gamma$. Here, $s$ is a section of $V$ and the maps $\Psi_g$ have been defined in (\ref{psig}). The prime on the right-hand-side of (\ref{equivcond}) indicates that an action of the induced equivariant structure on the co-tangent bundle should be included (since $\nabla(s)$ is a one-form).

What do the gauge fields associated to such an equivariant connection look like? To see this consider an open set $U_0$ which is sufficiently small that no two points inside it are mapped in to each other under the symmetry action $\Gamma$. We define the open sets $U_i=g_i(U_0)$ where $i=0,1,\ldots ,n$ which are the image of $U_0$ under the elements of the finite group. On each such patch $U_i$ we choose a frame $s_{i,a}$. This choice does not necessarily have to be aligned with the ``natural" choice $\Psi_{g_i}(s_{0,a})$ (which is swept out by the equivariant structure once a frame $s_{0,a}$ on $U_0$ has been fixed).
Let us parametrize the difference between those two frames by
\begin{equation} \label{apresequivcond}
 \Psi_{g_i}(s_{0,a})=P^b_{g_i,a}s_{i,b}\;.
\end{equation} 
It follows from the group law for $\Psi_g$ that the $P_{g}$ are matrices which must satisfy $P_{gh}=P_gP_h$. Consider a situation where we have sets of such matrices $P_g$, one set for all possible choices of initial open set $U_0$ within an open cover composed of such objects. Then, if we fix the frames $s_{i,a}$ once and for all, the sets of matrices $P_g$ encode the choice of equivariant structure on $V$. Call $A_i$ the gauge field for $\nabla$ on $U_i$ and relative to the frame $s_{i,a}$, that is,
\begin{equation} 
 \nabla(s_{i,a})=A^c_{ia}s_{i,c}\; .
\end{equation} 
Then, a short calculation shows that the equivariance condition (\ref{equivcond}) translates to the conditions
\begin{equation} \label{equivA}
 (g_i^{-1})^*A_0=P_i^{-1}A_iP_i+P_i^{-1}dP_i = A_{i,P_i}\; .
\end{equation} 
on the local gauge fields. This should be compared with (\ref{APdef}). In the special case when all $P_i=\mathbb{I}$ (which corresponds to a particular choice of equivariant structure), these conditions simplify to
\begin{equation}
  (g_i^{-1})^*A_0=A_i\; .
\end{equation}
In short, for two points on $X$ related by the symmetry, the corresponding gauge fields relate by a pull-back combined with a gauge transformation. This is certainly an intuitively reasonable definition for a gauge field we would expect to descend to a quotient. We discuss this further next. 

For later discussion it will be useful to note that it is obvious given the above definitions that any globally defined gauge transformation, $t:X\to G$ where $G$ is the gauge group, acting on an equivariant connection gives rise to another equivariant connection.\\[2mm]
{\bf Descent of connections:} Suppose we have a (local) section  $s$ of $V$ which is invariant, that is, $\Psi_g(s)=s$ and which descends to a section $\tilde{p}(s)$ of $\hat{V}$. For an equivariant connection $\nabla$ on $V$ such an invariant section satisfies, from (\ref{equivcond}), that
\begin{equation}
\Psi_g'(\nabla(s))=  \nabla(s)\; . \label{invcond}
\end{equation}
We therefore see that $\nabla(s)$ is an invariant section of $V\otimes TX^{\vee}$ and thus descends to a well-defined section of $\hat{V} \otimes T\hat{X}^{\vee}$ on the quotient. Thus we can define an associated downstairs connection by $\hat{\nabla}(\tilde{p}(s)):=\tilde{p}(\nabla (s))$, where by a slight abuse of notation we are using $\tilde{p}$ to indicate the descent of sections for two different sets of bundles.

To show this structure from a different perspective, let us start with a connection on the quotient and show that the pullback of the associated gauge field under the projection map, as defined in (\ref{APdef}), is an equivariant connection on $X$ as defined in (\ref{equivcond}). Let us denote by $\hat{\nabla}$ a connection on $\hat{X}$, and, given the projections $p_{i}:=p|_{U_{i}}:U_{i} \rightarrow  \hat{U}_{i} \subset \hat{X}$ of the open sets defined above (\ref{apresequivcond}), we can define the local pull-back connections $\nabla_i:=p_{i}^*\hat{\nabla}=d+A_{i}$. Here the associated gauge field $A_{i}$ is defined as the generalized pull-back of the gauge field on $\hat{X}$ following the discussion above, so that, according to (\ref{APdef})
\begin{equation} \label{equivcon}
p_{i}^*\hat{A}_i =  P_{i}^{-1}A_{i}P_{i}+P_{i}^{-1}dP_{i}\; ,
\end{equation} 
for some $P_{i}$. In this equation $A_i$ is the pullback of $\hat{A}_i$ under $p_{i}$ as an ordinary one-form.
Note that on any given patch we are free to choose the gauge transformation appearing in this expression to be $P=\mathbb{I}$, simply by performing a globally defined gauge transformation on each of the patches which agrees on all overlaps. If, for example, we choose $P_0$ to be the identity, then the gauge field $A_0=p_0^*\hat{A}$ on that patch is the straightforward pull-back of the downstairs gauge field. For a set of projections that are consistent with the group action we would have that $p_0 = p_i \circ g_i$ so that, in terms of ordinary pullbacks of one forms, $p_0^* = g_i^* \circ p_i^*$. Thus $p_0^*\hat{A} = g^* ( p_i^*( \hat{A}))$ so that $p_i^* \hat{A}=(g^{-1})^* A_0$. Given this and equation (\ref{equivcon}), we find the following relationship between the generalized pullbacks of the gauge fields to patches $U_0$ and $U_i$.
\begin{eqnarray} 
(g^{-1}_i)^* A_0 = P_{i}^{-1}A_{i}P_{i}+P_{i}^{-1}dP_{i}
\end{eqnarray}
This is precisely the relationship we saw in (\ref{equivA}) for an equivariant connection.
 \\[2mm]
\noindent{\bf Pullbacks and integration:} Let $f:X\rightarrow \hat{X}$ be a (smooth, surjective) map between manifolds, as before and $\nu$ a top form on $\hat{X}$. Then
\begin{equation} \label{pbint}
 \int_X f^*\nu={\rm deg}(f)\int_{\hat{X}}\nu\; ,
\end{equation}
where ${\rm deg}$ denotes the degree of a map. The degree is the integer which arises in the pullback of a top form. It is $+1$ for orientation-preserving diffeomorphisms and $-1$ for orientation-reversing diffeomorphisms. For non-injective maps which corresponds to an $N$-fold cover the degree is $\pm N$, with the sign is determined by what happens to the orientation \cite{bat}.

\vspace{0.2cm}

Note that this can immediately be applied to see how the holomorphic Chern-Simons invariant (\ref{mhcs}) descends to a quotient in some cases. We would define the Chern-Simons invariant in a heterotic string context by using a real isomorphism $f$ as in Section \ref{genproc}. We would choose this real isomorphism and equivariant structures on the two bundles involved to obey the commutativity condition (\ref{comfp}). In addition, it is necessary, if the integral (\ref{mhcs}) is to be well defined, that both connections are written with respect to the same local trivialization. With (\ref{comfp}) effectively saying that the two equivariant structures are the same for  the mapped connection $A$ and reference connection $A_0$, and with the choice of trivializations being the same, we then see from (\ref{apresequivcond}) that the gauge transformations appearing in the equivariance conditions for the two connections would be identical. Because the integrand in (\ref{mhcs}) is a gauge invariant if we transform both $A$ and $A_0$ simultaneously, we see that the group $\Gamma$ simply acts upon the integrand of the Chern-Simons invariant as though it were an ordinary differential form. Thus (\ref{pbint}) applies and we have that,
\begin{eqnarray} \label{upstairsdownstairs}
 \textnormal{CS}_{\hat{A}_0}(\hat{A})=\frac{1}{ \textnormal{deg}(p)}\textnormal{CS}_{A_0}(A) \;.
\end{eqnarray}
Here, $A$ and $A_0$ are the equivariant connections for the bundle $V\rightarrow X$ that are pullbacks of the connections $\hat{A}$ and $\hat{A}_0$ on $\hat{V}\rightarrow\hat{X}$, in the sense we have described above.



\begin{thebibliography}{99}
\ifx\doiref\asklfhas\newcommand{\doiref}[2]{\href{http://dx.doi.org/#1}{#2}}\fi
\raggedright 
\ifx\arxivref\asklfhas\newcommand{\arxivref}[2]{\href{http://arxiv.org/abs/#1}{arXiv:#1}}\fi
\raggedright

\bibitem{Candelas:1985en}
  P.~Candelas, G.~T.~Horowitz, A.~Strominger, E.~Witten,
  ``Vacuum Configurations for Superstrings,''
  \textsf{\doiref{10.1016/0550-3213(85)90602-9}{Nucl.\ Phys.\  {\bf B258 } (1985)  46-74}}.

\bibitem{gsw}  
  M.~B.~Green, J.~H.~Schwarz and E.~Witten,
  ``Superstring theory. Vol. 2: Loop amplitudes, anomalies and phenomenology,''
Cambridge University Press (1987),
\textsf{\doiref{10.1002/zamm.19880680631}{ https://doi.org/10.1002/zamm.19880680631}}.

\bibitem{Greene:1986bm}
  B.~R.~Greene, K.~H.~Kirklin, P.~J.~Miron and G.~G.~Ross,
  ``A Three Generation Superstring Model. 1. Compactification and Discrete Symmetries,''
 \textsf{\doiref{10.1016/0550-3213(86)90057-X}{Nucl.\ Phys.\ B {\bf 278} (1986) 667}}.

\bibitem{Greene:1986jb}
  B.~R.~Greene, K.~H.~Kirklin, P.~J.~Miron and G.~G.~Ross,
  ``A Three Generation Superstring Model. 2. Symmetry Breaking and the Low-Energy Theory,''
 \textsf{\doiref{10.1016/0550-3213(87)90662-6  }{Nucl.\ Phys.\ B {\bf 292} (1987) 606}}.

\bibitem{Braun:2011ni}
V.~Braun, P.~Candelas, R.~Davies and R.~Donagi,
``The MSSM Spectrum from (0,2)-Deformations of the Heterotic Standard Embedding,''
 \textsf{\doiref{doi:10.1007/JHEP05(2012)127}{JHEP \textbf{05} (2012), 127}, \arxivref{1112.1097}}.

\bibitem{Braun:2005ux}
V.~Braun, Y.~H.~He, B.~A.~Ovrut and T.~Pantev,
``A Heterotic standard model,''
 \textsf{\doiref{doi:10.1016/j.physletb.2005.05.007}{Phys. Lett. B \textbf{618} (2005), 252-258}, \arxivref{hep-th/0501070}}.

\bibitem{Braun:2005bw}
V.~Braun, Y.~H.~He, B.~A.~Ovrut and T.~Pantev,
``A Standard model from the E(8) x E(8) heterotic superstring,''
 \textsf{\doiref{doi:10.1088/1126-6708/2005/06/039}{JHEP \textbf{06} (2005), 039}, \arxivref{hep-th/0502155}}.

\bibitem{Braun:2005zv}
V.~Braun, Y.~H.~He, B.~A.~Ovrut and T.~Pantev,
``Vector bundle extensions, sheaf cohomology, and the heterotic standard model,''
Adv. Theor. Math. Phys. \textbf{10} (2006) no.4, 525-589
\textsf{\doiref{doi:10.4310/ATMP.2006.v10.n4.a3}{Adv. Theor. Math. Phys. \textbf{10} (2006) no.4, 525-589}, \arxivref{hep-th/0505041}}.
  
\bibitem{Bouchard:2005ag}
V.~Bouchard and R.~Donagi,
``An SU(5) heterotic standard model,''
\textsf{\doiref{doi:10.1016/j.physletb.2005.12.042}{Phys. Lett. B \textbf{633}, 783-791 (2006)}, \arxivref{hep-th/0512149}}.
  
  \bibitem{Anderson:2009mh}
L.~B.~Anderson, J.~Gray, Y.~H.~He and A.~Lukas,
``Exploring Positive Monad Bundles And A New Heterotic Standard Model,''
\textsf{\doiref{doi:10.1007/JHEP02(2010)054}{JHEP \textbf{02}, 054 (2010)}, \arxivref{0911.1569}}.
  
  \bibitem{Anderson:2011ns}
L.~B.~Anderson, J.~Gray, A.~Lukas and E.~Palti,
``Two Hundred Heterotic Standard Models on Smooth Calabi-Yau Threefolds,''
\textsf{\doiref{doi:10.1103/PhysRevD.84.106005}{Phys. Rev. D \textbf{84}, 106005 (2011)}, \arxivref{1106.4804}}.

\bibitem{Anderson:2012yf}
L.~B.~Anderson, J.~Gray, A.~Lukas and E.~Palti,
``Heterotic Line Bundle Standard Models,''
\textsf{\doiref{doi:10.1007/JHEP06(2012)113}{JHEP \textbf{06}, 113 (2012)}, \arxivref{1202.1757}}.

  \bibitem{Gray:2003vw}
J.~Gray and A.~Lukas,
``Gauge five-brane moduli in four-dimensional heterotic models,''
\textsf{\doiref{doi:10.1103/PhysRevD.70.086003}{Phys. Rev. D \textbf{70}, 086003 (2004)}, \arxivref{hep-th/0309096}}.

\bibitem{Candelas:2016usb}
P.~Candelas, X.~de la Ossa and J.~McOrist,
``A Metric for Heterotic Moduli,''
\textsf{\doiref{doi:10.1007/s00220-017-2978-7}{Commun. Math. Phys. \textbf{356} (2017) no.2, 567-612}, \arxivref{1605.05256}}.

\bibitem{McOrist:2016cfl}
J.~McOrist,
``On the Effective Field Theory of Heterotic Vacua,''
\textsf{\doiref{doi:10.1007/s11005-017-1025-0}{Lett. Math. Phys. \textbf{108} (2018) no.4, 1031-1081}, \arxivref{1606.05221}}.

\bibitem{Blesneag:2018ygh}
S.~Blesneag, E.~I.~Buchbinder, A.~Constantin, A.~Lukas and E.~Palti,
``Matter field Kähler metric in heterotic string theory from localisation,''
\textsf{\doiref{doi:10.1007/JHEP04(2018)139}{JHEP \textbf{04}, 139 (2018)}, \arxivref{1801.09645}}.

\bibitem{Candelas:2018lib}
P.~Candelas, X.~De La Ossa, J.~McOrist and R.~Sisca,
``The Universal Geometry of Heterotic Vacua,''
\textsf{\doiref{doi:10.1007/JHEP02(2019)038}{JHEP \textbf{02}, 038 (2019)}, \arxivref{1810.00879}}.
    
\bibitem{Gukov:1999ya}
S.~Gukov, C.~Vafa and E.~Witten,
``CFT's from Calabi-Yau four folds,''
Nucl. Phys. B \textbf{584}, 69-108 (2000)
doi:10.1016/S0550-3213(00)00373-4
[arXiv:hep-th/9906070 [hep-th]].
    
\bibitem{Conrad:2000tk}
J.~O.~Conrad,
``On fractional instanton numbers in six-dimensional heterotic E(8) x E(8) orbifolds,''
\textsf{\doiref{doi:10.1088/1126-6708/2000/11/022}{JHEP \textbf{11}, 022 (2000)}, \arxivref{hep-th/0009251}}.

\bibitem{Gukov:2003cy}
S.~Gukov, S.~Kachru, X.~Liu and L.~McAllister,
``Heterotic moduli stabilization with fractional Chern-Simons invariants,''
\textsf{\doiref{doi:10.1103/PhysRevD.69.086008}{Phys.\ Rev.\ D \textbf{69}, 086008 (2004)}, \arxivref{hep-th/0310159}}.

\bibitem{Apruzzi:2014dza}
F.~Apruzzi, F.~F.~Gautason, S.~Parameswaran and M.~Zagermann,
``Wilson lines and Chern-Simons flux in explicit heterotic Calabi-Yau compactifications,''
\textsf{\doiref{doi:10.1007/JHEP02(2015)183}{JHEP \textbf{02}, 183 (2015)}, \arxivref{1410.2603}}.

  \bibitem{Cicoli:2013rwa}
M.~Cicoli, S.~de Alwis and A.~Westphal,
``Heterotic Moduli Stabilisation,''
\textsf{\doiref{doi:10.1007/JHEP10(2013)199}{JHEP \textbf{10}, 199 (2013)}, \arxivref{1304.1809}}.

\bibitem{Jockers:2009ti}
H.~Jockers, P.~Mayr and J.~Walcher,
``On N=1 4d Effective Couplings for F-theory and Heterotic Vacua,''
\textsf{\doiref{doi:10.4310/ATMP.2010.v14.n5.a3}{Adv. Theor. Math. Phys. \textbf{14}, no.5, 1433-1514 (2010)}, \arxivref{0912.3265}}.

  \bibitem{Ashmore:2018ybe}
A.~Ashmore, X.~De La Ossa, R.~Minasian, C.~Strickland-Constable and E.~E.~Svanes,
``Finite deformations from a heterotic superpotential: holomorphic Chern-Simons and an $L_\infty$ algebra,''
\textsf{\doiref{doi:10.1007/JHEP10(2018)179}{JHEP \textbf{10}, 179 (2018)}, \arxivref{1806.08367}}.
   
\bibitem{Strominger:1986uh}
A.~Strominger,
``Superstrings with Torsion,''
\textsf{\doiref{doi:10.1016/0550-3213(86)90286-5}{Nucl. Phys. B \textbf{274}, 253 (1986)}}.

\bibitem{Hull:1986kz}
C.~Hull,
``Compactifications of the Heterotic Superstring,''
\textsf{\doiref{doi:10.1016/0370-2693(86)91393-6}{Phys. Lett. B \textbf{178}, 357-364 (1986)}}.

\bibitem{Rohm:1985jv}
R.~Rohm and E.~Witten,
``The Antisymmetric Tensor Field in Superstring Theory,''
\textsf{\doiref{doi:10.1016/0003-4916(86)90099-0}{Annals Phys.\  \textbf{170}, 454 (1986)}}.

\bibitem{Lukas:1997rb}
A.~Lukas, B.~A.~Ovrut and D.~Waldram,
``Gaugino condensation in M theory on s**1 / Z(2),''
\textsf{\doiref{doi:10.1103/PhysRevD.57.7529}{Phys.\ Rev.\ D \textbf{57}, 7529-7538 (1998)}, \arxivref{hep-th/9711197}}.
   
\bibitem{Freed:1986zx}
D.~Freed,
``Determinants, Torsion, and Strings,''
Commun. Math. Phys. \textbf{107}, 483-513 (1986)
doi:10.1007/BF01221001

\bibitem{Dijkgraaf:1989pz}
R.~Dijkgraaf and E.~Witten,
``Topological Gauge Theories and Group Cohomology,''
Commun. Math. Phys. \textbf{129}, 393 (1990)
doi:10.1007/BF02096988

\bibitem{Freed:2008jq}
D.~S.~Freed,
``Remarks on Chern-Simons Theory,''
[arXiv:0808.2507 [math.AT]].
   
\bibitem{Thomas:1998uj}
R.~Thomas,
``A Holomorphic Casson invariant for Calabi-Yau three folds, and bundles on K3 fibrations,''
J. Diff. Geom. \textbf{54}, no.2, 367-438 (2000),
\textsf{\arxivref{math/9806111}}.

\bibitem{donaldsonbook}
S.~K.~Donaldson,
``Floer Homology Groups in Yang-Mills Theory,"
Cambridge University Press (2002),
\textsf{\doiref{doi:10.1017/CBO9780511543098}{https://doi.org/10.1017/CBO9780511543098}}.

\bibitem{thomasthesis}
R.~Thomas,
``Gauge Theory on Calabi-Yau Manifolds,"
PhD Thesis, University of Oxford, (1997)

\bibitem{Witten:1985bt}
E.~Witten,
``Topological Tools in Ten-dimensional Physics,''
\textsf{\doiref{doi:10.1142/S0217751X86000034}{Int.\ J.\ Mod.\ Phys.\ A \textbf{1}, 39 (1986)}}.

\bibitem{Wen:1985qj}
X.~Wen and E.~Witten,
``Electric and Magnetic Charges in Superstring Models,''
\textsf{\doiref{doi:10.1016/0550-3213(85)90592-9}{Nucl.\ Phys.\ B \textbf{261}, 651-677 (1985)}}.

\bibitem{Witten:1985mj}
E.~Witten,
``Global Anomalies in String Theory,''
Print-85-0620 (PRINCETON).

\bibitem{Witten:1999eg}
E.~Witten,
``World sheet corrections via D instantons,''
\textsf{\doiref{doi:10.1088/1126-6708/2000/02/030}{JHEP \textbf{02}, 030 (2000)}, \arxivref{hep-th/9907041}}.

\bibitem{yau}
S.-T.~Yau,
``On the Ricci curvature of a compact K\"ahler manifold and the complex monge-amp\'ere equation, I"
\textsf{\doiref{doi:10.1002/cpa.3160310304}{Comm. Pure Appl. Math., 31: 339-411, (1978)}}.

\bibitem{duy1}
K.~Uhlenbeck and S.-T.~Yau,
``On the existence of Hermitian Yang-Mills connections in stable bundles", 
\textsf{\doiref{10.1002/cpa.3160390714}{Comm. Pure App. Math., {\bf 39}, 257, (1986)}}.

\bibitem{duy2}
S.~Donaldson,
``Anti Self-Dual Yang-Mills Connections over Complex Algebraic Surfaces and Stable Vector Bundles",
\textsf{\doiref{10.1112/plms/s3-50.1.1}{Proc. London Math. Soc., {\bf 3}, 1, (1985)}}.

\bibitem{Witten:1985xc}
E.~Witten,
``Symmetry Breaking Patterns in Superstring Models,''
\textsf{\doiref{doi:10.1016/0550-3213(85)90603-0}{Nucl. Phys. B \textbf{258}, 75 (1985)}}.

\bibitem{Strominger:1985it}
A.~Strominger and E.~Witten,
``New Manifolds for Superstring Compactification,''
\textsf{\doiref{doi:10.1007/BF01216094}{Commun. Math. Phys. \textbf{101}, 341 (1985)}}.

\bibitem{Candelas:1987se}
P.~Candelas,
``Yukawa Couplings Between (2,1) Forms,''
\textsf{\doiref{doi:10.1016/0550-3213(88)90351-3}{Nucl. Phys. B \textbf{298}, 458 (1988)}}.

\bibitem{Candelas:1987kf}
P.~Candelas, A.~Dale, C.~Lutken and R.~Schimmrigk,
``Complete Intersection Calabi-Yau Manifolds,''
\textsf{\doiref{doi:10.1016/0550-3213(88)90352-5}{Nucl. Phys. B \textbf{298}, 493 (1988)}}.

\bibitem{Candelas:1990rm}
P.~Candelas, X.~C.~De La Ossa, P.~S.~Green and L.~Parkes,
``A Pair of Calabi-Yau manifolds as an exactly soluble superconformal theory,''
\textsf{\doiref{doi:10.1016/0550-3213(91)90292-6}{AMS/IP Stud. Adv. Math. \textbf{9}, 31-95 (1998)}}.

\bibitem{Berglund:1993ax}
P.~Berglund, P.~Candelas, X.~De La Ossa, A.~Font, T.~Hubsch, D.~Jancic and F.~Quevedo,
``Periods for Calabi-Yau and Landau-Ginzburg vacua,''
\textsf{\doiref{doi:10.1016/0550-3213(94)90047-7}{Nucl. Phys. B \textbf{419}, 352-403 (1994)}, \arxivref{hep-th/9308005}}.

\bibitem{Donaldson} 
S.~K.~Donaldson, 
``Some numerical results in complex differential geometry,"
\textsf{\doiref{doi:10.4310/PAMQ.2009.v5.n2.a2}{Pure\ Appl.\ Math.\ Q.\ {\bf 5}, no. 2, part 1 571-618 (2009)}, \arxivref{math/0512625}}.
  
 \bibitem{Headrick:2005ch} 
  M.~Headrick and T.~Wiseman,
  ``Numerical Ricci-flat metrics on K3,''
 \textsf{\doiref{doi:10.1088/0264-9381/22/23/002}{ Class.\ Quant.\ Grav.\  {\bf 22}, 4931 (2005)}, \arxivref{hep-th/0506129}}.
  
 \bibitem{Douglas:2006hz} 
  M.~R.~Douglas, R.~L.~Karp, S.~Lukic and R.~Reinbacher,
  ``Numerical solution to the hermitian Yang-Mills equation on the Fermat quintic,''
 \textsf{\doiref{doi:10.1088/1126-6708/2007/12/083}{JHEP {\bf 0712}, 083 (2007)}, \arxivref{hep-th/0606261}}.
  
 \bibitem{Douglas:2006rr} 
  M.~R.~Douglas, R.~L.~Karp, S.~Lukic and R.~Reinbacher,
  ``Numerical Calabi-Yau metrics,''
 \textsf{\doiref{doi:10.1063/1.2888403}{ J.\ Math.\ Phys.\  {\bf 49}, 032302 (2008)}, \arxivref{hep-th/0612075}}.
 
 \bibitem{Braun:2007sn} 
  V.~Braun, T.~Brelidze, M.~R.~Douglas and B.~A.~Ovrut,
  ``Calabi-Yau Metrics for Quotients and Complete Intersections,''
\textsf{\doiref{doi:10.1088/1126-6708/2008/05/080}{ JHEP {\bf 0805}, 080 (2008)}, \arxivref{0712.3563}}.
  
  \bibitem{Braun:2008jp} 
  V.~Braun, T.~Brelidze, M.~R.~Douglas and B.~A.~Ovrut,
  ``Eigenvalues and Eigenfunctions of the Scalar Laplace Operator on Calabi-Yau Manifolds,''
 \textsf{\doiref{doi:10.1088/1126-6708/2008/07/120}{JHEP {\bf 0807}, 120 (2008)}, \arxivref{0805.3689}}.
  
  \bibitem{Headrick:2009jz} 
  M.~Headrick and A.~Nassar,
  ``Energy functionals for Calabi-Yau metrics,''
  \textsf{\doiref{doi:10.4310/ATMP.2013.v17.n5.a1}{Adv.\ Theor.\ Math.\ Phys.\  {\bf 17}, no. 5, 867 (2013)}, \arxivref{0908.2635}}.

\bibitem{Anderson:2010ke} 
  L.~B.~Anderson, V.~Braun, R.~L.~Karp and B.~A.~Ovrut,
  ``Numerical Hermitian Yang-Mills Connections and Vector Bundle Stability in Heterotic Theories,''
 \textsf{\doiref{doi:10.1007/JHEP06(2010)107}{ JHEP {\bf 1006}, 107 (2010)}, \arxivref{1004.4399}}.
  
  \bibitem{Anderson:2011ed} 
  L.~B.~Anderson, V.~Braun and B.~A.~Ovrut,
  ``Numerical Hermitian Yang-Mills Connections and Kahler Cone Substructure,''
  \textsf{\doiref{doi:10.1007/JHEP01(2012)014}{JHEP {\bf 1201}, 014 (2012)}, \arxivref{1103.3041}}.
 
\bibitem{Ashmore:2019wzb} 
  A.~Ashmore, Y.~H.~He and B.~A.~Ovrut,
  ``Machine learning Calabi-Yau metrics,''
  \textsf{\arxivref{1910.08605}}.

\bibitem{Cui:2019uhy}
W.~Cui and J.~Gray,
``Numerical Metrics, Curvature Expansions and Calabi-Yau Manifolds,''
\textsf{\doiref{doi:10.1007/JHEP05(2020)044}{JHEP \textbf{05} (2020), 044}, \arxivref{1912.11068}}.
 
\bibitem{Hubsch:1986ny} T.~H\"ubsch,
  ``Calabi-Yau Manifolds: Motivations and Constructions,''
  \textsf{\doiref{10.1007/BF01210616}{Commun.\ Math.\ Phys.\  {\bf 108} (1987) 291}}.

\bibitem{Green:1986ck} P.~Green and T.~H\"ubsch,
  ``Calabi-Yau Manifolds as Complete Intersections in Products of Complex Projective Spaces,''
  \textsf{\doiref{10.1007/BF01205673}{Commun.\ Math.\ Phys.\  {\bf 109} (1987) 99}}.

\bibitem{Candelas:1987du} P.~Candelas, C.~A.~Lutken and R.~Schimmrigk,
  ``Complete Intersection Calabi-Yau Manifolds. 2. Three Generation Manifolds,''
  \textsf{\doiref{10.1016/0550-3213(88)90173-3}{Nucl.\ Phys.\ B {\bf 306} (1988) 113}}.

\bibitem{Anderson:2015iia}
L.~B.~Anderson, F.~Apruzzi, X.~Gao, J.~Gray and S.~Lee,
``A new construction of Calabi-Yau manifolds: Generalized CICYs,''
\textsf{\doiref{doi:10.1016/j.nuclphysb.2016.03.016}{Nucl. Phys. B \textbf{906}, 441-496 (2016)}, \arxivref{1507.03235}}.

\bibitem{Berglund:2016yqo}
P.~Berglund and T.~H\"ubsch,
``On Calabi-Yau generalized complete intersections from Hirzebruch varieties and novel $K3$-fibrations,''
\textsf{\doiref{doi:10.4310/ATMP.2018.v22.n2.a1}{Adv. Theor. Math. Phys. \textbf{22}, 261-303 (2018)}, \arxivref{1606.07420}}.

\bibitem{Berglund:2016nvh}
P.~Berglund and T.~Hubsch,
``A Generalized Construction of Calabi-Yau Models and Mirror Symmetry,''
\textsf{\doiref{doi:10.21468/SciPostPhys.4.2.009}{SciPost Phys. \textbf{4}, no.2, 009 (2018)}, \arxivref{1611.10300}}.

\bibitem{Kreuzer:2000xy}
M.~Kreuzer and H.~Skarke,
``Complete classification of reflexive polyhedra in four-dimensions,''
\textsf{\doiref{doi:10.4310/ATMP.2000.v4.n6.a2}{Adv. Theor. Math. Phys. \textbf{4} (2002), 1209-1230}, \arxivref{hep-th/0002240}}.

\bibitem{Kreuzer:2002uu} M.~Kreuzer and H.~Skarke,
  ``PALP: A Package for analyzing lattice polytopes with applications to toric geometry,''
  \textsf{\doiref{10.1016/S0010-4655(03)00491-0}{Comput.\ Phys.\ Commun.\  {\bf 157} (2004) 87}, \arxivref{math/0204356}}.

\bibitem{Altman:2014bfa}
R.~Altman, J.~Gray, Y.~He, V.~Jejjala and B.~D.~Nelson,
``A Calabi-Yau Database: Threefolds Constructed from the Kreuzer-Skarke List,''
 \textsf{\doiref{doi:10.1007/JHEP02(2015)158}{JHEP \textbf{02}, 158 (2015)}, \arxivref{1411.1418}}.

\bibitem{Braun:2010vc}
V.~Braun,
``On Free Quotients of Complete Intersection Calabi-Yau Manifolds,''
 \textsf{\doiref{doi:10.1007/JHEP04(2011)005}{JHEP \textbf{04}, 005 (2011)}, \arxivref{1003.3235}}.
 
\bibitem{Candelas:2008wb}
P.~Candelas and R.~Davies,
``New Calabi-Yau Manifolds with Small Hodge Numbers,''
 \textsf{\doiref{doi:10.1002/prop.200900105}{Fortsch. Phys. \textbf{58} (2010), 383-466}, \arxivref{0809.4681}}.

\bibitem{Candelas:2010ve}
P.~Candelas and A.~Constantin,
``Completing the Web of $Z_3$ - Quotients of Complete Intersection Calabi-Yau Manifolds,''
 \textsf{\doiref{doi:10.1002/prop.201200044}{Fortsch. Phys. \textbf{60} (2012), 345-369}, \arxivref{1010.1878}}.
 
\bibitem{Candelas:2015amz}
P.~Candelas, A.~Constantin and C.~Mishra,
``Hodge Numbers for CICYs with Symmetries of Order Divisible by 4,''
 \textsf{\doiref{doi:10.1002/prop.201600005}{Fortsch. Phys. \textbf{64} (2016) no.6-7, 463-509}, \arxivref{1511.01103}}.

\bibitem{Candelas:2016fdy}
P.~Candelas, A.~Constantin and C.~Mishra,
``Calabi?Yau Threefolds with Small Hodge Numbers,''
 \textsf{\doiref{doi:10.1002/prop.201800029}{Fortsch. Phys. \textbf{66} (2018) no.6, 1800029}, \arxivref{1602.06303}}.

\bibitem{Constantin:2016xlj} 
  A.~Constantin, J.~Gray and A.~Lukas,
  ``Hodge Numbers for All CICY Quotients,''
\textsf{\doiref{doi:10.1007/JHEP01(2017)001}{ JHEP {\bf 1701}, 001 (2017)}, \arxivref{1607.01830}}.

\bibitem{Candelas:2017ive}
P.~Candelas and C.~Mishra,
``Highly Symmetric Quintic Quotients,''
\textsf{\doiref{doi:10.1002/prop.201800017}{Fortsch. Phys. \textbf{66}, no.4, 1800017 (2018)}, \arxivref{1709.01081}}.

\bibitem{Anderson:2007nc}
L.~B.~Anderson, Y.~H.~He and A.~Lukas,
``Heterotic Compactification, An Algorithmic Approach,''
\textsf{\doiref{10.1088/1126-6708/2007/07/049}{JHEP \textbf{07}, 049 (2007)}, \arxivref{hep-th/0702210}}.

\bibitem{Anderson:2011ty}
L.~B.~Anderson, J.~Gray, A.~Lukas and B.~Ovrut,
``The Atiyah Class and Complex Structure Stabilization in Heterotic Calabi-Yau Compactifications,''
\textsf{\doiref{doi:10.1007/JHEP10(2011)032}{JHEP \textbf{10}, 032 (2011)}, \arxivref{1107.5076}}.

\bibitem{cicylist}
A machine readable version of the CICY list and the symmetries used in this paper can be found \href{http://www-thphys.physics.ox.ac.uk/projects/CalabiYau/CicyQuotients/Cicy_Quotients/Cicy_Quotients.html}{here}. (http://www-thphys.physics.ox.ac.uk/projects/CalabiYau/CicyQuotients/\\Cicy\_Quotients/Cicy\_Quotients.html)

\bibitem{kirkklassen}
P.~A.~Kirk and E.~P.~Klassen,
``Chern-Simons Invariants of 3-Manifolds and Representation Spaces of Knot Groups,"
\textsf{\doiref{doi:10.1007/BF01446898}{Math. Ann. 287 (1990) 343-367}}.

\bibitem{Rozansky:1993zx}
L.~Rozansky,
``A Large k asymptotics of Witten's invariant of Seifert manifolds,''
\textsf{\doiref{doi:10.1007/BF02099272}{Commun. Math. Phys. \textbf{171}, 279-322 (1995)}, \arxivref{hep-th/9303099}}.

\bibitem{Nishi}
H.~Nishi,
``SU(n) Chern-Simons Invariants of Seifert Fibered 3-Manifolds,"
\textsf{\doiref{doi:10.1142/S0129167X98000130}{Int. J. Math. 09 (1998) 295}}.


\bibitem{clemens_thomas}
H.~Clemens (with an appendix by R.~Thomas),
``Cohomology and Obstructions II: Curves on K-Trivial Threefolds,"
\textsf{\arxivref{math/0206219}}.

\bibitem{Anderson:2010mh}
L.~B.~Anderson, J.~Gray, A.~Lukas and B.~Ovrut,
``Stabilizing the Complex Structure in Heterotic Calabi-Yau Vacua,''
\textsf{\doiref{doi:10.1007/JHEP02(2011)088}{JHEP \textbf{02}, 088 (2011)}, \arxivref{1010.0255}}.

\bibitem{Anderson:2013qca}
L.~B.~Anderson, J.~Gray, A.~Lukas and B.~Ovrut,
``Vacuum Varieties, Holomorphic Bundles and Complex Structure Stabilization in Heterotic Theories,''
\textsf{\doiref{doi:10.1007/JHEP07(2013)017}{JHEP \textbf{07}, 017 (2013)}, \arxivref{1304.2704}}.

\bibitem{Anderson:2011cza}
L.~B.~Anderson, J.~Gray, A.~Lukas and B.~Ovrut,
``Stabilizing All Geometric Moduli in Heterotic Calabi-Yau Vacua,''
\textsf{\doiref{doi:10.1103/PhysRevD.83.106011}{Phys. Rev. D \textbf{83}, 106011 (2011)}, \arxivref{1102.0011}}.


\bibitem{okonek}
C.~Okonek, M.~Schneider and H.~Spindler,
``Vector Bundles on Complex Projective Spaces,"
Springer, 1980,
\textsf{\doiref{doi:10.1007/978-1-4757-1460-9}{https://doi.org/10.1007/978-1-4757-1460-9}}.

\bibitem{Douglas:2004yv}
M.~R.~Douglas and C.~g.~Zhou,
``Chirality change in string theory,''
\textsf{\doiref{doi:10.1088/1126-6708/2004/06/014}{JHEP \textbf{06}, 014 (2004)}, \arxivref{hep-th/0403018}}.

\bibitem{huybrechts}
D.~Huybrechts,
``Complex Geometry, An Introduction,"
Springer, 2005,
\textsf{\doiref{doi:10.1007/b137952}{https://doi.org/10.1007/b137952}}.

\bibitem{bradlow_schaposnik}
S.~B.~Bradlow and L.~P.~Schaposnik,
``Higgs bundles and exceptional isogenies,"
Res. Math. Sci. {\bf 3}, 14 (2016). \textsf{\doiref{doi.org/10.1186/s40687-016-0062-0}{https://doi.org/10.1186/s40687-016-0062-0}}.

\bibitem{bat}
R.~Bott and L.~W.~Tu,
``Differential Forms in Algebraic Topology,"
Springer, 1982,
\textsf{\doiref{doi:10.1007/978-1-4757-3951-0}{https://doi.org/10.1007/978-1-4757-3951-0}}.



\end{thebibliography}
\end{document}